\documentclass[aps,prx,twocolumn,longbibliography,superscriptaddress]{revtex4-2}

\usepackage{dcolumn}
\usepackage{bm}
\usepackage[english]{babel}
\usepackage{lipsum} 
\usepackage{times}
\usepackage{amsfonts,amssymb,bm}
\usepackage{epsfig}
\usepackage{mathrsfs}
\usepackage{color,soul}
\usepackage[normalem]{ulem}
\usepackage{bbold}
\usepackage{amssymb}
\usepackage{pstricks}
\usepackage{multirow}
\usepackage{graphicx}
\usepackage{textcomp}
\usepackage{braket}
\usepackage{empheq}
\usepackage{hyperref}
\hypersetup{colorlinks,%
citecolor=red,%
filecolor=black,%
linkcolor=blue,%
urlcolor=blue}
\usepackage{color}
\usepackage{rotating}
\usepackage{amsthm}
\usepackage{siunitx}
\usepackage{amsmath}
\usepackage{eurosym}
\usepackage[babel]{csquotes}

\usepackage[font=small]{caption}
\usepackage{mwe}
\DeclareMathOperator{\tr}{Tr}

\usepackage[framemethod=default]{mdframed}
\usepackage[labelfont=bf,labelsep=quad,justification=raggedright]{caption}
\usepackage{newfloat}

\DeclareFloatingEnvironment[name={S}]{suppfigure}
\DeclareFloatingEnvironment[name={S}]{supptable}
\allowdisplaybreaks

\definecolor{bazaar}{rgb}{0.6, 0.47, 0.48}
\definecolor{green1}{rgb}{0.33, 0.7, 0.69}

\newmdenv[linecolor=white,backgroundcolor=gray!15!]{myframe}

\AtBeginDocument{%
    \newwrite\bibnotes
    \def\bibnotesext{Notes.bib}
    \immediate\openout\bibnotes=\jobname\bibnotesext
    \immediate\write\bibnotes{@CONTROL{REVTEX41Control}}
    \immediate\write\bibnotes{@CONTROL{%
    apsrev41Control,author="08",editor="1",pages="1",title="0",year="1"}}
     \if@filesw
     \immediate\write\@auxout{\string\citation{apsrev41Control}}%
    \fi
}%

\begin{document}

\title{Quantum superposition of thermodynamic evolutions with opposing time's arrows}

\author{Giulia Rubino}
\thanks{Corresponding author: giulia.rubino@univie.ac.at}
\affiliation{Vienna Center for Quantum Science and Technology (VCQ), Faculty of Physics, University of Vienna, Boltzmanngasse 5, 1090, Vienna, Austria}
\affiliation{Quantum Engineering Technology Labs, H. H. Wills Physics
Laboratory and Department of Electrical \& Electronic Engineering,
University of Bristol, Bristol BS8 1FD, United Kingdom}

\author{Gonzalo Manzano}
\affiliation{Institute for Cross-Disciplinary Physics and Complex Systems (IFISC) UIB-CSIC, Campus Universitat Illes Balears, E-07122 Palma de Mallorca, Spain}
\affiliation{Institute for Quantum Optics and Quantum Information (IQOQI), Austrian Academy of Sciences, Boltzmanngasse 3, 1090 Vienna, Austria}

\author{\v{C}aslav Brukner}
\affiliation{Vienna Center for Quantum Science and Technology (VCQ), Faculty of Physics, University of Vienna, Boltzmanngasse 5, 1090, Vienna, Austria}
\affiliation{Institute for Quantum Optics and Quantum Information (IQOQI), Austrian Academy of Sciences, Boltzmanngasse 3, 1090 Vienna, Austria}


\date{\today}
\begin{abstract}
Microscopic physical laws are time-symmetric, hence, a priori there exists no preferential temporal direction. However, the second law of thermodynamics allows one to associate the ``forward'' temporal direction to a positive variation of the total entropy produced in a thermodynamic process, and a negative variation with its ``time-reversal'' counterpart.
This definition of a temporal axis is normally considered to apply in both classical and quantum contexts. Yet, quantum physics admits also superpositions between forward and time-reversal processes, whereby the thermodynamic arrow of time becomes quantum-mechanically undefined. In this work, we demonstrate that a definite thermodynamic time's arrow can be restored by a quantum measurement of entropy production, which effectively projects such superpositions onto the forward (time-reversal) time-direction when large positive (negative) values are measured.
Remarkably, for small values (of the order of plus or minus one), the amplitudes of forward and time-reversal processes can interfere, giving rise to entropy-production distributions featuring a more or less reversible process than either of the two components individually, or any classical mixture thereof.
\end{abstract}
\maketitle


\section{Introduction}

In spite of it being seemingly straightforward, physics is still nowadays seeking to provide a comprehensive understanding of the apparent passage of time \cite{Halliwell_1996}. The concept of time flow is intimately related to the observation of a change in physical systems. However, the recognition that, at their most fundamental level, physical systems generally obey time-reversible laws led to the realisation that systems' evolutions do not intrinsically differentiate between forward and backward time directions. Attempts to uphold with physical arguments the evidence of the time flow are being made on multiple fronts, mainly on the basis of empirical observations: we see that entropy in the universe increases (thermodynamic time's arrow), that the universe expands (cosmological time's arrow), that causes always precede their effects (causal time's arrow).
Likewise, there have been several proposals as to the explanation of the time's arrow in a quantum-mechanical contexts~\cite{PhysRevLett.103.080401, PhysRevLett.104.148901, PhysRevE.81.061130, PhysRevE.89.052102, PhysRevX.7.031022}. The peculiarity of the quantum framework is that it enables for processes to be placed in quantum superposition. Applied to the notion of thermodynamic time's arrow, this implies that quantum mechanics can allow the superposition of thermodynamic processes (namely, dynamic processes wherein a system of interest exchanges either heat, work, or both with other systems, the environment and/or external agents) producing opposite variations in the entropy.
This raises the question of how a well-defined thermodynamic arrow of time can be established in the quantum framework when such superpositions are in place.
To address this question, in this work we show that a measurement of the entropy production has a decisive role in restoring a definite thermodynamic time's arrow, 
and we investigate interference effects in such superpositions. 
Our investigations bear a conceptual similarity to the field of indefinite quantum causality, wherein the order of operations is placed in a quantum superposition~\cite{Chiribella2012, Chiribella2013, Oreshkov2012}. Note, however, that there is a crucial difference between these two types of studies. In indefinite quantum causality, operations are performed in the same temporal direction (here referred to as ``forward'') in each amplitude of the superposition. In contrast, in the present case we analyse superpositions of thermodynamic processes with opposing thermodynamic arrows of time.

In thermodynamics, the time's arrow is introduced 
by the second law of thermodynamics, according to which the total entropy of the universe 
can only either increase, or remain constant. Consequently, one might think that observations of entropy changes are all we need to distinguish the past from the future: an overall increase in entropy shall be identified with the direction of time ``forward'', while a overall decrease in entropy with its ``time-reversal'' counterpart. Yet, for a microscopic system, fluctuations blur the direction of the time's arrow, and the time flow is only defined \textit{on average}. More specifically, in this regime the time's arrow cannot be inferred, as both positive and negative entropy changes can be observed with comparable probability in a single experimental run. As a consequence, for such systems the two opposite time's arrows become classically indistinguishable. The extension of this indistinguishability to the quantum domain gives rise to quantum superpositions between opposite time's arrows, whose investigation is the focus of the present work.

In what follows, we will explore how a definite time's arrow arises in quantum superpositions between ``forward'' and ``time-reversal'' processes (i.e., thermodynamic processes whose quenches are related by time-inversion symmetry).
We will start by constructing a quantum superposition between two such processes (Section \ref{sec:superp_of_dynamics}), and the mathematical framework for their evaluation (Section \ref{sec:ExtTPM}). Then, we will show that quantum measurements of the dissipative work $W_\text{diss}$ (or, equivalently, entropy production $\Delta S_{\mathrm{tot}}$) 
can restore the time directionality of the process. 
The dissipative work $W_\text{diss} = W - \Delta F$ is the amount of work $W$ invested in a thermodynamic transformation between equilibrium states having a free energy difference $\Delta F$, which cannot be recovered by reversing the process. Furthermore, the relation between the dissipative work and the entropy production (or total entropy) $\Delta S_\mathrm{tot}$ in the process is established through the relation: $\Delta S_{\mathrm{tot}} = \beta \, W_\text{diss}$, where $\beta = (k_B T)^{-1}$ is the inverse temperature, with $k_B$ being the Boltzmann constant and $T$ the temperature of the bath~\cite{Kawai:2007,Parrondo:2009,Serra:2015}. 
In Section \ref{sec:Effect_Projection} we will show that, when the measured dissipative work equals $\beta W_{\text{diss}} \gg 1$, the superposition is effectively projected onto the forward process, whereas when $\beta W_{\text{diss}} \ll - 1$, it is effectively projected onto the time-reversal one, hence recovering 
a definite thermodynamic arrow of time (albeit, in each individual execution of the experiment, the outcome ``forward" or ``time reversal" is random). 
Conversely, 
when $\beta \vert W_\text{diss} \vert$ is of the order of one, the forward and the time-reversal thermodynamic processes can quantum mechanically interfere under certain conditions, 
resulting in a work probability distribution describing work fluctuations which have no classical counterpart. More precisely, in the case of interference, the probabilities take on values which cannot be obtained by any classical (convex) mixture of the forward and the time-reversal processes (Section \ref{sec:interference}). 

\section{Results}

\subsection{Superposition of forward and time-reversal dynamics} \label{sec:superp_of_dynamics}

We start by defining the framework used to characterize thermodynamic processes and work fluctuations. First, we will introduce all the necessary elements to formally construct a state representing the quantum superposition of a thermodynamic process evolving in the forward temporal direction, and one evolving in the opposite (time-reversal) direction. Then, we will discuss how to characterize work and entropy production fluctuations in such superposition states using an extended two-point-measurement (TPM) scheme, and we illustrate how the outcomes achieved through processes with well-defined time directions can be recovered inside our framework.

We consider a thermodynamic system $S$ being, in both forward and time-reversal processes, initially in equilibrium with a thermal reservoir at inverse temperature $\beta$. The process occurring in the forward direction will be realized by a quench $U(t,0)$ induced by the time-dependent Hamiltonian $H\bigl(\lambda(t)\bigr)$ executing a controlled protocol $\Lambda \equiv \{\lambda(t) ; 0 \leq t \leq \tau \}$ in the time-frame $t \in [0, \tau]$, followed by a final thermalisation in contact with the reservoir at $\beta$. Here, $U(t_1, t_2) = \mathbf{\overrightarrow{T}} \mathrm{exp}\bigl[ -\frac{i}{\hbar} \int_{t_1}^{t_2} d\nu \, H\bigl(\lambda(\nu)\bigr)\bigr]$, where $\mathbf{\overrightarrow{T}}$ is the so-called ``time-ordering'' operator resulting from the Dyson decomposition.
Its time-reversal twin will be described by a quench $\tilde{U}(\tau -t, 0)$ associated to the implementation of the operational time-reversal protocol $\tilde{\Lambda} \equiv \{ {\lambda}(\tau - t) ; 0 \leq t \leq \tau \}$, where in both cases $\lambda$ is a control parameter, and again the quench is followed by a final thermalisation step. 
The micro-reversibility principle for non-autonomous systems establishes a  strong relation between forward and time-reversal quenches lying at the core of fluctuation theorems~\cite{Campisi:2011,Gaspard:2008}: 
 \begin{equation} \label{eqn:microrev}
\tilde{U}(\tau-t,0) = \Theta \, U^\dagger(\tau,t)  \, \Theta^\dagger,
\end{equation}
where $\Theta$ denotes the (anti-unitary) time-reversal operator acting on the system's Hilbert space, which flips the sign of observables with odd parity under time-reversal. This operator verifies the relations $\Theta \, \, \mathbb{1}i = - \mathbb{1}i \, \Theta$, and $\Theta \, \Theta^\dagger = \Theta^\dagger \, \Theta = \mathbb{1}$. 

In order to describe superpositions of forward and time-reversal processes, the initial equilibrium states of the system $S$ can be purified by including some environmental degrees of freedom $E$ with a generic Hamiltonian $H_E$ in the description. 
These purifications are not unique, and they can be represented by joint states of the system and the environment of the form
\begin{subequations}\label{eqn:pur_state}
\begin{align}
\label{eqn:pur_state_0}
& \vert \psi _0 \rangle_{S,E}  = \sum_{k} \sqrt{\frac{e^{-\beta E^{(0)}_k}}{Z_{0}}} \, \vert E_k^{(0)} \rangle_S \, \vert \varepsilon_k^{(0)} \rangle_E,\\
\label{eqn:pur_state_tau}
& \vert \tilde{\psi}_0 \rangle_{S,E}  = \sum_{k} \sqrt{\frac{e^{-\beta E^{(\tau)}_k}}{Z_{\tau}}} \, \Theta \vert E_k^{(\tau)}  \rangle_S \, \vert \varepsilon_k^{(\tau)} \rangle_E,
\end{align}
\end{subequations}
where $E_k^{(0)}$ and $E_k^{(\tau)}$ are the eigenvalues of the Hamiltonian at times $t = \{0, \tau \}$, i.e., $H[\lambda(0)]$ and $H[\lambda(\tau)]$, whereas $\vert E_k^{(0)} \rangle_S$ and $\vert E_k^{(\tau)} \rangle_S$ are the corresponding eigenvectors (for the sake of brevity, we will henceforth omit the subscript $S$ in the system's energy eigenvectors). Furthermore, $\vert \varepsilon_k^{(0)} \rangle_E$, $\vert \varepsilon_k^{(\tau)} \rangle_E$ represent the corresponding sets of states of the environmental degree of freedom, 
which can always be chosen as sets of orthogonal states. Notice that the environment may possess further degrees of freedom which are not entangled with the system under consideration, and which we will thus not explicitly account for.

The state $\vert \psi_0 \rangle_{S,E}$ above corresponds to the initial state of the process evolving in the forward direction as defined by $\Lambda$, whereas  $\vert \tilde{\psi}_0 \rangle_{S,E}$ is the initial state of the time-reversed process as defined by $\tilde{\Lambda}$. 
Notice that, by tracing out the environmental degrees of freedom, we recover the corresponding Gibbs thermal states for the system $\rho_0^{\text{th}} \equiv \mathrm{Tr}_E \bigl(\vert \psi_0 \rangle \langle \psi_0 \vert_{S,E}\bigr) = e^{-\beta H[\lambda(0)]}/Z_0$ and $\tilde{\rho}_0^{\text{th}} \equiv \mathrm{Tr}_E \bigl(\vert \tilde{\psi}_0 \rangle \langle \tilde{\psi}_0 \vert_{S,E} \bigr) = \Theta \, e^{-\beta H[\lambda(\tau)]} \Theta^\dagger /Z_\tau$, being $Z_{0} = \mathrm{Tr} \bigl(e^{-\beta H[\lambda(0)]}\bigr)$, and $Z_{\tau} = \mathrm{Tr} \bigl(e^{-\beta H[\lambda(\tau)]}\bigr)$ the partition functions.

Moreover, we introduce an auxiliary system $A$ whose two orthogonal states $\lbrace \vert 0\rangle_A, \vert 1\rangle_A\rbrace$ govern the evolution of the process in the two temporal directions. This is a quantum analogue of the coin tossed to decide classically which process to run (forward or time reversal). 
With this in place, the global Hamiltonian of the system, the environment, and the auxiliary qubit reads $\mathcal{H}(t) \equiv \big(\ket{0}\bra{0}_A \otimes H[\lambda(t)] + \ket{1}\bra{1}_A \otimes \Theta {H}[\lambda(\tau-t)] \Theta^\dagger \big) \otimes \mathbb{1}_{E} +  \mathbb{1}_{S,A} \otimes H_E$.
We then entangle each orthogonal auxiliary state to one of the initial states in Eq.~$\eqref{eqn:pur_state}$. The overall initial state of thermodynamic system, environment and auxiliary system reads therefore:
\begin{equation}
\label{eqn:initial_state}
\vert \Psi_0 \rangle_{S,E,A} = \alpha_0  \, \vert \psi_0 \rangle_{S,E}\otimes\vert 0 \rangle_{A} + \alpha_1  \, \vert {\tilde{\psi}}_0 \rangle_{S,E}\otimes\vert 1 \rangle_{A},
\end{equation}
with arbitrary coefficients $\alpha_0, \alpha_1 \in \mathbb{C}$, $\vert\alpha_0\vert^2 + \vert\alpha_1\vert^2=1$.
If, subsequently, in each branch of the superposition in Eq.~$\eqref{eqn:initial_state}$ the forward and time-reversal quenches are respectively applied, the evolved state at some arbitrary instant of time $t \in [0, \tau]$ is given by $\vert \Psi (t) \rangle_{S,E,A} = \alpha_0  \, \bigl[ U(t,0) \otimes \mathbb{1}_{E,A} \bigr] \vert \psi_0 \rangle_{S,E} \otimes \vert 0 \rangle_{A} + \alpha_1  \vert \bigl[ \tilde{U}(t,0) \otimes \mathbb{1}_{E,A} \bigr] \vert \tilde{\psi}_0 \rangle_{S,E} \otimes \,\vert 1 \rangle_{A}$. In this expression, the first and the second amplitudes correspond to the forward and the time-reversal processes, 
respectively. Furthermore, we assume that the system does not interact with the environment 
during the timescale of the quenches 
(however, after the quench, the system thermalises through the interaction with the thermal reservoir). This is verified whenever the quenches are implemented in a fast timescale as compared to the characteristic relaxation time of the system in interaction with the environment~\cite{Dorner:2012}, or when the system is artificially disconnected from the environment during the quench implementation and reconnected after it. 
Furthermore, we will consider the quenches $U(t,0)$ and $\tilde{U}(t,0)$ in the superposition to be implemented by some  external (classical) control. As we discuss in the Supplementary Note II~\cite{SM}, this limit is adequate in our setup, and it corresponds to the case in which the control mechanism acts approximately as an ideal reservoir of energy and coherence~\cite{PhysRevLett.113.150402, Malabarba:2015, Korzekwa:2016, Bartlett:2007}, as is the case, for instance, with lasers or radio-frequency pulses.


Taking a gas enclosed in a vessel as a pictorial example, the aforementioned state can be constructed by entangling the position of the piston with a further auxiliary quantum system, thereby establishing a quantum superposition of the following two processes: \textit{i.} a process wherein the gas particles are initially in thermal equilibrium confined in one half of the vessel by a piston, and the piston is pulled outwards, and \textit{ii.} the reverse process, in which the piston is pushed towards the gas, starting from an initial state where the gas occupies the entire vessel in thermal equilibrium.

\subsection{Extended two-point measurement scheme}
\label{sec:ExtTPM}

We will now measure the work of the system undergoing the above-mentioned superposition of forward and time-reversal dynamics. In order to implement such a measurement, we formally construct a procedure described by a set of measurement operators forming a completely positive and trace-preserving (CPTP) map. In this regard, we will refer to a standard TPM procedure to measure work in quantum thermodynamic processes~\cite{Campisi:2011}.
Implementations of the TPM in quantum setups~\cite{Dorner:2013,Mazzola:2013,Serra:2014,Roncaglia:2014,deChiara:2015}, as well as suitable extensions~\cite{Perarnau:2017, Aberg:2018, Debarba:2019, Mohammady:2019}, 
have recently received increasing attention. Our procedure can be seen as a generalization of the TPM scheme to situations where different thermodynamic processes are allowed to be superposed, and can consequently interfere.

In the TPM scheme, work is defined as the energy difference between the initial and final states of the system, which are measured through ideal projective measurements of the system Hamiltonian implemented before and after the thermodynamic process associated to the protocol $\Lambda$~\cite{Talkner:2007,Talkner:2016}. This measurement scheme can be performed, individually, both for the forward and the time-reversal processes, enabling the construction of the work probability distributions $P(W)$ and $\tilde{P}(W)$, respectively.

As far as the forward process is concerned, the probability to observe a transition $\ket{E_n^{(0)}} \rightarrow \ket{E_m^{(\tau)}}$ is given by $p_{n,m} = p_{m\vert n} \, p_n^{(0)}$,
where $p_n^{(0)} = e^{-\beta E_n^{(0)}}/Z_0$ is the probability of observing the energy $E_n^{(0)}$ at $t = 0$, and $p_{m\vert n} = \left \vert \langle E_m^{(\tau)} \vert \, U(\tau, 0)\vert E_n^{(0)} \rangle \right \vert^2$ is the conditional probability of measuring $E_m^{(\tau)}$ at $t=\tau$ after having measured $E_n^{(0)}$ at the beginning of the process. Similarly, for the time-reversal process one has $\tilde{p}_{m,n} = \tilde{p}_{n\vert m} \, \tilde{p}_m^{(0)}$, 
where $\tilde{p}_m^{(0)} = e^{-\beta E_m^{(\tau)}}/Z_\tau$ is the probability to obtain the energy $E_m^{(\tau)}$ at the beginning of the time-reversal process, and $\tilde{p}_{n | m} = \left \vert \langle {E}_n^{(0)} \vert \Theta^\dagger \tilde{U}(\tau, 0) \Theta \vert {E}_m^{(\tau)} \rangle \right \vert^2$ is the corresponding conditional probability for observing the inverse transition $\Theta \ket{E_m^{(\tau)}} \rightarrow \Theta \ket{E_n^{(0)}}$ given that one obtained $E_m^{(\tau)}$ in the first measurement. The micro-reversibility principle in Eq.~\eqref{eqn:microrev} relates the conditional probabilities in the forward and time-reversal processes as $\tilde{p}_{n\vert m} = p_{m\vert n}$ \cite{Campisi:2011,Gaspard:2008}. 

The TPM scheme allows one to compute the stochastic work invested by the external driver in a single realisation of the protocol $\Lambda$, $W_{n, m} \equiv E_m^{(\tau)} - E_n^{(0)}$, associated to the outcomes of initial and final energy measurements. Its probability distribution reads:
\begin{equation}\label{eqn:wprob}
P(W) = \sum_{n, m} p_{n, m} \cdot \delta(W- W_{n, m}).      
\end{equation}
Analogously, the probability distribution associated to the work invested in the time-reversal protocol,  $\tilde{W}_{n, m}= E_n^{(0)} - E_m^{(\tau)} = - W_{n, m}$, is given by:
\begin{equation}\label{eqn:wprobr}
\tilde{P}(W) = \sum_{n, m} \tilde{p}_{n, m} \cdot \delta(W - \tilde{W}_{n, m}).          
\end{equation}

\begin{figure}[tb]
\centering
\includegraphics[width=\columnwidth]{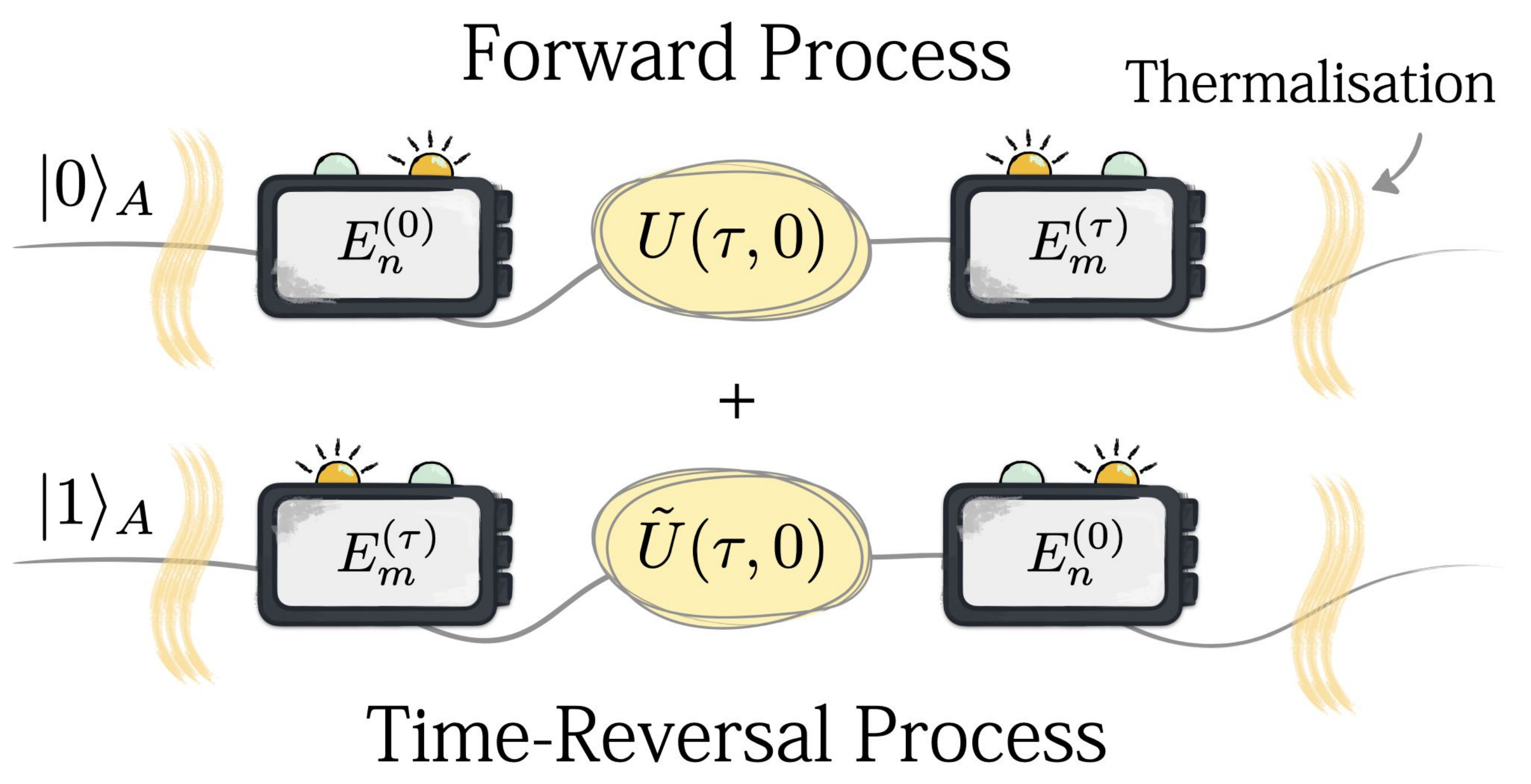}
\captionof{figure}{\footnotesize \textbf{Schematic representation of a superposition of a forward thermodynamic quench with its time-reversal counterpart.} A thermodynamic system $S$ and its environment are coupled to an auxiliary system $A$ in a suitable entangled state. Depending on the state of the auxiliary system, $\vert 0\rangle_A$ or $\vert 1\rangle_A$, when the state of the environment is traced out, the system $S$ is initially prepared in a thermal state of the initial or final Hamiltonians, $H(0)$ and $H(\tau)$, respectively. This is then sent  through a thermodynamic quench $U(t,0)$ or its time reversal $\tilde{U}(t,0)$ in the time-frame $t \in [0,\tau]$. Before and after each quench, the system's energy is measured. The measurement outcomes $E_n^{(0)}$ and $E_m^{(\tau)}$ are found when the auxiliary system is in $\vert 0\rangle_A$, whereas the outcomes $E_m^{(0)}$ and $E_n^{(\tau)}$ are obtained when the auxiliary system is in $\vert 1\rangle_A$. 
After these measurements, the system may eventually undergo a second thermalisation with the environment. 
Note that the first (second) measurement when the auxiliary system is in $|0\rangle_A$, and the second (first) measurement when it is in $|1\rangle_A$ are physically one and the same measurement. A possible implementation of this scheme is reported in Fig.~\ref{img:sketch_extended}.} 
\label{img:sketch}
\end{figure}

Hereafter, we consider an extension of the TPM scheme in which we include energy measurements at $t=0$ and $t=\tau$ in both branches of the superposition between a forward and a time-reversal processes, as illustrated in Fig.~\ref{img:sketch}. More precisely, starting with the initial state in Eq.~\eqref{eqn:initial_state}, and conditionally on the auxiliary  state, we consider the application of the projectors $\ket{E_n^{(0)}}\bra{E_n^{(0)}}$ and $\Theta \ket{E_m^{(\tau)}}\bra{E_m^{(\tau)}} \Theta^\dagger$ to the initial states $\ket{\psi_0}_{S,E}$ and $\ket{\tilde{\psi}_0}_{S,E}$, respectively. Subsequently, the unitary quenches $U(\tau, 0)$ and $\tilde{U}(\tau, 0)$ are implemented in each branch, after which the projectors $\ket{E_m^{(\tau)}}\bra{E_m^{(\tau)}}$ and $\Theta \ket{E_n^{(0)}}\bra{E_n^{(0)}} \Theta^\dagger$ are respectively applied. Consequently, given the outcomes $E_n
^{(0)}$ and $E_m^{(\tau)}$, a work $W_{n,m}$ is invested in the forward-dynamics branch by applying the protocol $\Lambda$, whereas the work invested in its time-reversal counterpart $\tilde{\Lambda}$ is $\tilde{W}_{n, m} = -W_{n, m}$ (that is, the same amount of work as in the forward dynamics is here extracted). 

The operator representing the application of the scheme through which the work $W_{n,m}$ is obtained can be written as: 
\begin{align}\label{eqn:M_W}
&M_{{n,m}} = 
\vert E_m^{(\tau)} \rangle \langle E_m^{(\tau)} \vert U(\tau, 0)  \vert E_n^{(0)} \rangle \langle E_n^{(0)}\vert \otimes \mathbb{1}_E \otimes \vert 0 \rangle \langle 0 \vert_A \nonumber \\ 
&+\Theta \vert {E}_n^{(0)} \rangle \langle {E}_n^{(0)} \vert \Theta^\dagger \tilde{U}(\tau, 0) \Theta \vert {E}_m^{(\tau)} \rangle \langle {E}_m^{(\tau)} \vert \Theta^\dagger \otimes \mathbb{1}_E \otimes  \vert 1 \rangle \langle 1 \vert_A 
\end{align}
The set of operators $\{ M_{n,m}\}$ forms a CPTP map, 
$\mathcal{E}(\rho) \equiv \sum_{n,m} M_{n,m} \rho M_{n,m}^\dagger$, acting on the composite system $S,E,A$ and fulfilling $\sum_{n,m} M_{n,m}^\dagger M_{n,m} = \mathbb{1}_{S,E,A}$. The map $\mathcal{E}$ describes the average effect of the measurement scheme on an arbitrary initial state of the composite system $\rho$, while the operations $\mathcal{E}_{n,m} (\rho) \equiv M_{n,m} \rho M_{n,m}^\dagger$ provide
the probability $\mathcal{P}(W) \equiv \sum_{n,m} \tr[\mathcal{E}_{n,m} (\rho)] \delta \bigl(W - W_{n,m} \bigr)$ to measure the work $W$.

\begin{figure}[tb]
\centering
\includegraphics[width=\columnwidth]{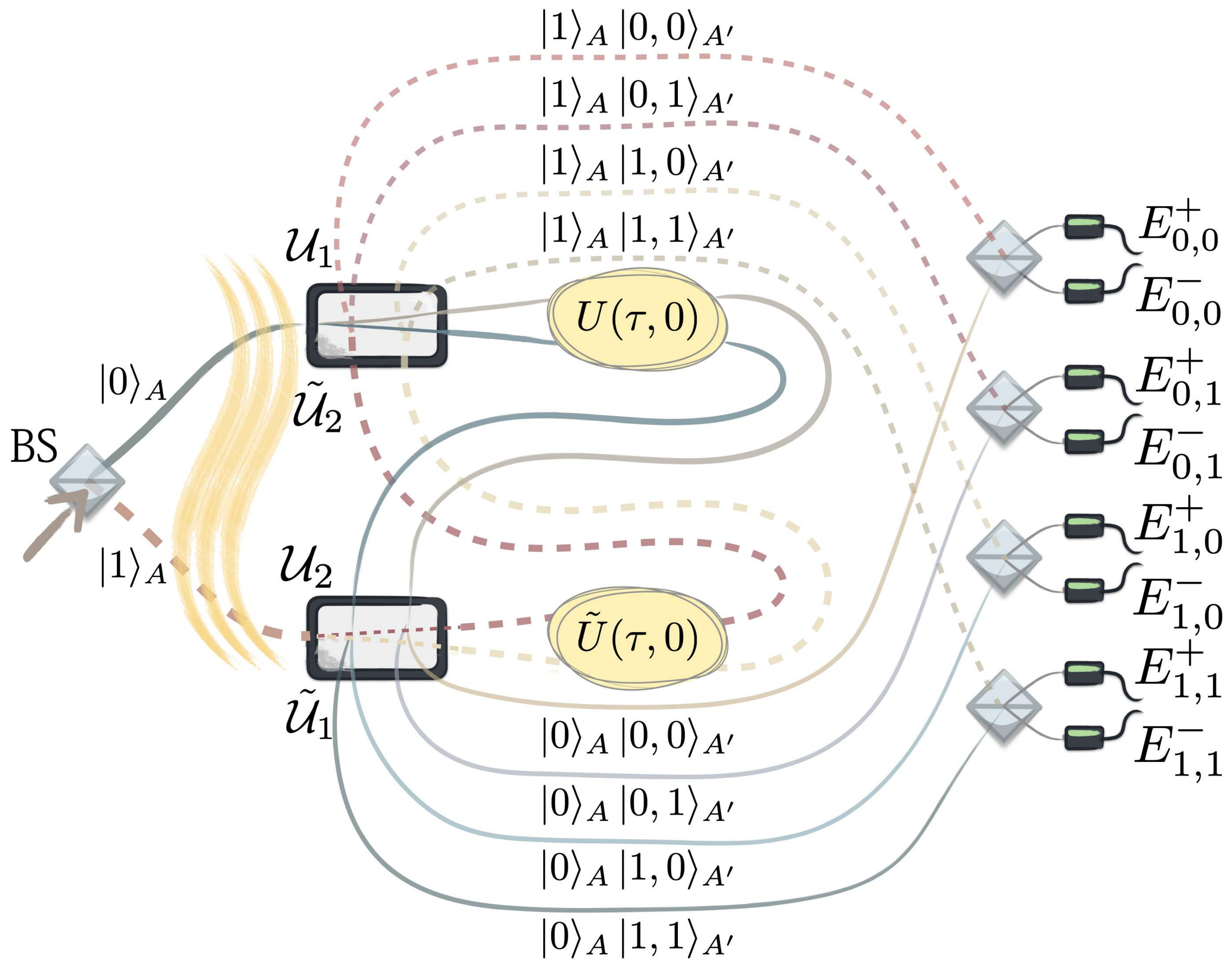}
\captionof{figure}{\footnotesize {\textbf{Sketch of a possible implementation of the extended TPM scheme in the case of binary results from each of the two measurements.} A first beam splitter (BS) creates a quantum superposition of the auxiliary state in $\vert 0 \rangle_A$, $\vert 1 \rangle_A$ as represented by the upper (solid) and lower (dashed) paths in the left part of the figure, over which the initial state in Eq.~\eqref{eqn:initial_state} is prepared. In order not to reveal the which-path information, projective measurements of the system energy are replaced by unitary operators $\mathcal{U}_{1,2}$ and $\tilde{\mathcal{U}}_{1,2}$, coupling the system with two additional internal registers $A'$, which are initially prepared in the state $|0,0\rangle_{A'}$. Encoding pairs of system energy eigenstates $(n,m)$ onto the registers $A'$ leads to further subdivisions into different paths (middle part of the figure), which are recombined and measured only at the final stage of the interferometric scheme. 
When the auxiliary system is in the state $\vert 0 \rangle_A$ ($\vert 1 \rangle_A$), the thermodynamic system is first subjected to a unitary $\mathcal{U}_1$ ($\mathcal{\tilde{U}}_1$), then to the thermodynamic process $U(\tau,0)$ [$\tilde{U}(\tau,0)$] within the time interval $[0,\tau$], and finally to a second unitary $\mathcal{U}_2$ ($\mathcal{\tilde{U}}_2$), see solid (dashed) paths in the figure. 
Unitary $\mathcal{U}_1$ encodes the energy of the eigenstates $|E_n^{(0)}\rangle$ of the thermodynamic system into the first register $|n,0\rangle_{A'}$ ($n=0,1$) of the auxiliary system, while unitary $\mathcal{U}_2$ encodes the energy of the eigenstates $|E_m^{(\tau)}\rangle$ into the states $|n,m\rangle_{A'}$ ($m=0,1$) of the second register. The four possible outcomes are indicated as four solid paths (bottom part of the figure), each labeled as $|0\rangle_A \vert n,m\rangle_{A'}$. Similarly, the unitaries $\mathcal{\tilde{U}}_1$ and $\mathcal{\tilde{U}}_2$ encode the energies of the thermodynamic system before and after the quench in the second and the first register respectively, when the auxiliary system is in the state $|1\rangle_A$. This is represented by the four dashed paths labeled as $|1\rangle_A \vert m,n\rangle_{A'}$ (top part of the figure). The coherence of the auxiliary system's states is thus maintained until the end of the interferometer, as depicted in the rightmost part of the figure. There, the states $|0\rangle_A |n,m\rangle_{A'}$ and $|1\rangle|m,n\rangle_{A'}$ for each pair $n,m$ are interfered pairwise through further BSs, and finally measured (thereby revealing the system energy values). The results of the final measurements over the auxiliary degree of freedom $A$ in the diagonal basis $\{ \ket{\pm}_A \}$ corresponding to pairs $n,m$, are indicated by the symbols $E_{n,m}^\pm$.
The unitaries $\mathcal{U}_{1,2}$ and $\tilde{\mathcal{U}}_{1,2}$, together with the final measurement after recombining paths, replace 
the initial and final energy measurements of $E_n^{(0)}$ and $E_m^{(\tau)}$ in the TPM scheme in Fig.~\ref{img:sketch}.}} 
\label{img:sketch_extended}
\end{figure}

It is important to stress that the operations $\mathcal{E}_W$ preserve the coherence between the forward and time-reversal thermodynamic processes. Indeed, performing a standard quantum measurement on the process would destroy the coherence, as it would reveal the time at which the measurement has been performed, and, from this, also whether the outcome $E_m$ was observed before (in the forward process) or after the outcome $E_n$ (in the time-reversal process). In other words, such a measurement would reveal the time direction, and it would be equivalent to the measurement of the auxiliary  qubit in the basis $\bigl\{\vert 0 \rangle_A, \vert 1 \rangle_A\bigr\}$. However, there exist also measurement schemes in which the result is encoded in an auxiliary system through its entanglement with the measured system, and the result is then read only at the end of the whole evolution, thereby preserving its coherence. (Such a measurement scheme was recently used to measure the system undergoing superposition of causal orders~\cite{Rubinoe1602589}.) In such a scheme, the system on which the thermodynamic quenches act and the auxiliary system can be encoded on two different degrees of freedom of the same quantum system. If the auxiliary degree of freedom is of sufficient dimension, it is possible to encode the results of each measurement taking place within the process in a state of this system. More precisely, suppose that the auxiliary system has two additional registers $A'$ which can store the results of the two energy measurements. When the auxiliary system is in the $|0\rangle_A$ ($|1\rangle_A$) state, the thermodynamic system is subject to an unitary ${\cal U}_1$ (${\cal \tilde{U}}_1$) that couples the energy of the system to the first (second) register of the auxiliary system. This results in an overall unitary that entangles the thermodynamic system with the auxiliary system:
\begin{equation}
    |0\rangle_A\langle 0| \otimes {\cal U}_1 + |1\rangle_A\langle 1| \otimes {\cal \tilde{U}}_1,
\end{equation}
where
\begin{align}
&   {\cal U}_1 |E^{(0)}_n\rangle_S |x,y\rangle_{A'} = |E^{(0)}_n\rangle_S |x\oplus n,y\rangle_{A'}, \\
 & {\cal \tilde{U}}_1 \Theta| E^{(\tau)}_m\rangle_S |x,y\rangle_{A'} = \Theta| E^{(\tau)}_m\rangle_S |x,y \oplus m\rangle_{A'},
\end{align}
for any basis state $|x,y\rangle_{A'}$ of the two registers. Here, the symbol $\oplus$ means the sum modulo the total number of different energy values. Subsequently, the  thermodynamic system is subject to a quench, followed by another entangling unitary 
\begin{equation}
    |0\rangle_A\langle 0| \otimes {\cal U}_2 + |1\rangle_A\langle 1| \otimes {\cal \tilde{U}}_2,
\end{equation}
with 
\begin{align}
    &   {\cal U}_2 | E^{(\tau)}_m\rangle_S |x,y\rangle_{A'} = | E^{(\tau)}_m\rangle_S |x,y \oplus m\rangle_{A'}, \\
 & {\cal \tilde{U}}_2 \Theta |E^{(0)}_n\rangle_S |x,y\rangle_{A'} = \Theta |E^{(0)}_n\rangle_S |x \otimes n,y \rangle_{A'},
\end{align}
which now couples the energy of the thermodynamic system after the quench in the second (first) register when the auxiliary system is in the state $|0\rangle_A$ ($|1\rangle_A$). 
If the two registers are initially prepared in the state $|0,0\rangle_{A'}$, their final state $|n,m\rangle_{A'}$ will encode both energy values.
The coherence of the overall state has to be maintained until the end of the entire thermodynamic process when the auxiliary system is  measured in the basis $\{(|0\rangle_A \pm |1\rangle_A)/\sqrt{2} \}$ to erase any information as to whether the system has gone through the  ``forward" or ``time-reversal" process (which might be encoded, for instance, in the temporal or directional mode of the auxiliary system). A sketch of a possible experimental realisation of the extended TPM scheme is shown in Fig.~\ref{img:sketch_extended} in the case of two measurements outcomes for $E_n^{(0)}$, $E_m^{(\tau)}$. For simplicity, in this study we consider only two states of the auxiliary system (Fig.~\ref{img:sketch}). Nevertheless, all the conclusions drawn herein can be extended to the case of more than two states.

In order to evaluate the work probability distribution in the extended TMP scheme, 
it is also crucial to take into account the mutual phases between the conditional probabilities. We thus write, in general
\begin{subequations}
\begin{align}
& \langle E_m^{(\tau)} \vert \, U(\tau, 0) \vert E_n^{(0)} \rangle := \sqrt{p_{m \vert n}} \, e^{i \Phi_{n,m}},\\
& \langle E_n^{(0)} \vert \, U^\dagger(\tau, 0) \vert E_m^{(\tau)} \rangle := \sqrt{\tilde{p}_{n \vert m}} \, e^{- i \tilde{\Phi}_{m,n}},
\end{align}
\end{subequations}
and we notice that
\begin{align}
\sqrt{\tilde{p}_{n\vert m}} e^{- i \tilde{\Phi}_{m,n}} &= \langle E_n^{(0)} \vert \, U^\dagger(\tau, 0) \vert E_m^{(\tau)} \rangle \notag\\
&= \bigl( \sqrt{p_{m \vert n}} \, e^{i \Phi_{n,m}} \bigr)^\ast = \sqrt{p_{m\vert n}} \, e^{- i \Phi_{n,m}}, \notag
\end{align}
from which we get $\Phi_{n,m} = \tilde{\Phi}_{m,n}$, since $\tilde{p}_{n\vert m} = p_{m\vert n}$.

We now consider the concatenation of the operation $M_{n,m}$ with a projection of the auxiliary  qubit onto an arbitrary state $|\xi \rangle_A$. 
By applying this 
sequence of operations to the initial state in Eq.~\eqref{eqn:initial_state}, we derive the (unnormalized) state of the composite system associated to the work outcome $W_{n,m}$ and projection of the auxiliary  qubit onto $\ket{\xi}_A$: 
\begin{align}
\label{eqn:M_DSE_Psi0}
\vert \Psi_{n,m}^\xi \rangle_{S,E,A} &\equiv \bigl( \mathbb{1}_{S,E} \,\otimes |\xi\rangle \langle\xi|_A\bigr) \circ M_{n,m} \vert \Psi_0 \rangle_{S,E,A} \notag\\
&= \ket{\Xi_0^\xi} +  \ket{\Xi_1^\xi},
\end{align}
where we identified the two branches of the superposition corresponding to the forward $(\ket{\Xi_0^\xi})$ and the time-reversal dynamics $(\ket{\Xi_1^\xi})$. They read, respectively: 
\begin{subequations}\label{eqn:pieces}
\begin{align}
\ket{\Xi_0^\xi} =&  \alpha_0 \langle \xi \vert 0 \rangle 
\sqrt{p_{n,m}} \, e^{i \Phi_{n,m}}  \, \vert E_m^{(\tau)} \rangle \, \vert \varepsilon^{(0)}_n \rangle_E |\xi\rangle_A \\ 
\ket{\Xi_1^\xi} =& \alpha_1 \langle \xi \vert 1 \rangle 
\sqrt{p_{n,m}} \, e^{-\frac{\beta}{2} (W_{n, m} - \Delta F) - i \Phi_{n,m}} \nonumber \\ & \Theta \vert {E}_n^{(0)} \rangle \, \vert  {\varepsilon}^{(\tau)}_m \rangle_E |\xi\rangle_A 
\end{align}
\end{subequations}
where, in the second equation, we made use of $\tilde{p}_{n ,m} = p_{n, m} \,e^{-\beta (W_{n, m} - \Delta F)}$ (see the Supplementary Note III~\cite{SM}), and of the relation between the forward and time-reversal phases $\tilde{\Phi}_{m,n} = \Phi_{n,m}$. The final thermalization step, which effectively leads to the irreversible dissipation of work  $W_\mathrm{diss}$, occurs only after the projection onto the auxiliary qubit, and is thus not included within the (extended) TPM scheme.

The joint probability of measuring the work $W$ and projecting the auxiliary  state onto $\ket{\xi}_A$ is given by $\mathcal{P}(\xi,W) = \sum_{n,m} \bigl\vert \bigl\vert \vert \Psi_{n,m}^\xi \rangle_{S,E,A} \bigl\vert\bigl\vert^2 \delta(W - W_{n,m})$.
Furthermore, from the joint probabilities $\mathcal{P}(\xi,W)$, one can obtain the conditional ones $\mathcal{P}_\xi(W):= \mathcal{P}(W\vert\xi) = \mathcal{P}(\xi, W)/\mathcal{P}(\xi)$, which we will hereafter refer to as ``post-selected work probability distributions'', and where $\mathcal{P}(\xi) = \int dW \, \mathcal{P}(\xi, W)$. By introducing the notation $q^\xi_0 = \vert \alpha_0 \vert^2 \bigl\vert\langle \xi \vert 0 \rangle\bigr\vert^2/\mathcal{P}(\xi)$ and
$q^\xi_1 = \vert \alpha_1 \vert^2 \bigl\vert\langle \xi \vert 1 \rangle\bigr\vert^2/\mathcal{P}(\xi)$, 
we can rewrite $\mathcal{P}_\xi(W)$ as:
\begin{align} \label{eqn:wdist}
&\mathcal{P}_\xi(W) = 
q^\xi_0~ P(W) + q^\xi_1~\tilde{P}(-W) + 2\,\mathbb{Re}\bigl(I_\xi (W)\bigr),
\end{align}
where we identified the probability distributions for the work in the forward process $P(W)$, and in the time-reversal one $\tilde{P}(-W)$ as given in Eqs.~\eqref{eqn:wprob}-\eqref{eqn:wprobr}, respectively. From this, we obtain the interference term:
\begin{align} \label{eqn:Iw}
& I_{\xi}(W) = 
\dfrac{\alpha_0^\ast \alpha_1 \langle 0 | \xi \rangle \langle \xi | 1 \rangle}{\mathcal{P}(\xi)} 
\sum_{n, m} p_{n,m} 
 e^{-\frac{\beta}{2}(W_{n,m} - \Delta F)} ~\notag \\ 
& e^{-2i \Phi_{n,m}} \bra{E_m^{(\tau)}}\Theta \ket{E_{n}^{(0)}} \braket{\varepsilon_n^{(0)} | \varepsilon_{m}^{(\tau)}} \cdot \delta(W - W_{n, m}) 
\end{align}
The functional dependence of $\mathcal{P}_\xi(W)$ on $W$ consists of two parts: \textit{i.} an ``incoherent'' part, reflecting the fact that each work value $W$ obtained in the scheme is compatible with running the process in one or the other temporal direction with a given probability (i.e., investing the work $W$ when running the protocol $\Lambda$, and extracting the same amount of work $-W$ when executing its time-reversal counterpart $\tilde{\Lambda}$), and \textit{ii.} a ``coherent'' part, which is a genuinely quantum feature arising from the superposition of the two temporal directions of the quench.


In the case $\vert \alpha_0 \vert = \vert \alpha_1 \vert = 1/\sqrt{2}$, the forward state $\ket{\Xi_0^\xi}$ and the time-reversal one $\ket{\Xi_1^\xi}$ in Eq.~\eqref{eqn:pieces} have the same amplitudes in the superposition. Nevertheless, as in the standard scenario of well-defined temporal directions~\cite{SM}, one may use the properties of the work probability distribution $\mathcal{P}_\xi(W)$ together with Bayesian reasoning to infer the time's arrow of the thermodynamic process. As we will see shortly, in some cases, the thermodynamic time's arrow can be determined even in a single realisation of the process, which effectively projects the state $\vert \Psi_{n,m}^\xi \rangle_{S,E,A}$ onto either its forward or its time-reversal component.

\subsection{Effective projection onto a definite time's arrow}
\label{sec:Effect_Projection}

In the following, we demonstrate that measuring work values such that $W - \Delta F \gg \beta^{-1}$, or $W - \Delta F \ll - \beta^{-1}$, in single realisations of the extended TPM scheme effectively results in projecting the state $\vert \Psi_{m,n}^\xi \rangle_{S,E,A}$ in Eq.~\eqref{eqn:M_DSE_Psi0} onto either the forward or the time-reversal components in Eq.~\eqref{eqn:pieces} (i.e., $\ket{\Xi_0^\xi}$ or $\ket{\Xi_1^\xi}$, respectively). 
In order to show this, we consider the probabilities for the superposition state $\vert \Psi_{n,m}^\xi \rangle_{S,E,A}$ to be found in either $\vert\vert\ket{\Xi_0^\xi}\vert\vert^2$ or $\vert\vert\ket{\Xi_1^\xi}\vert\vert^2$, respectively. In particular, we notice that the term $\vert\vert \ket{\Xi_1^\xi}\vert\vert^2$ is upper bounded by
\begin{align}
\vert\vert \ket{\Xi_1^\xi} \vert\vert^2 &= \vert \alpha_1 \vert^2 \bigl\vert\langle \xi \vert 1 \rangle\bigr\vert^2 
p_{n,m} \, e^{- \beta (W_{n, m} - \Delta F)} \notag\\
& 
\leqslant e^{- \beta W_{\text{diss}}} \sum_{n, m}  \, p_{n,m} = e^{- \beta W_{\text{diss}}},
\end{align}
where we used the fact that $\vert \alpha_1 \vert^2 \bigl\vert\langle \xi \vert 1 \rangle\bigr\vert^2 \leqslant 1$, and $\sum_{n, m}  \, p_{n,m} = 1$. 
Consequently, in the limit $\beta W_{\text{diss}} \gg 1$, we have $\vert \vert \ket{\Xi_1^\xi} \vert \vert^2 \approx 0$, and hence $\vert \vert \ket{\Xi_0^\xi} \vert \vert^2 \approx 1$, that is, $\vert \Psi_{n,m}^\xi \rangle_{S,E,A} \simeq \ket{\Xi_0^\xi}$. Indeed, applying the detailed fluctuation theorem in Eq.~\eqref{eqn:fluct} to Eq.~\eqref{eqn:wdist}, we obtain:
\begin{align}
\label{eqn:M_W-on-Psi0}
\mathcal{P}_\xi(W) &= P(W) \bigl(q^\xi_0 + q^\xi_1 e^{- \beta W_\mathrm{diss}} \bigl)+ 2\,\mathbb{Re}\bigl(I_\xi (W)\bigr) \notag \\ & \approx  q^\xi_0 ~ P(W),
\end{align}
where we made use of the fact that $ I_\xi (W) \propto e^{- \beta W_{\text{diss}}/2}$.
Therefore, we obtained that, whenever one performs a measurement of the work in the extended TPM scheme and observes $W - \Delta F \gg \beta^{-1}$ (or, equivalently, $\Delta S = \beta W_{\text{diss}} \gg 1$), the state of the system is projected onto the forward component of the quantum superposition without measuring the auxiliary  qubit (similarly to what one would obtain, had one projected the joint state $\vert \Psi (t) \rangle_{S,E,A}$ through a projective measurement $\vert 0 \rangle \langle 0\vert_A$ on the auxiliary  system, and subsequently observed the work value $W$). The probability to observe this work value in the extended TPM scheme is given by Eq.~\eqref{eqn:M_W-on-Psi0}.

Analogously, whenever the result of the extended TPM scheme is such that $W - \Delta F \ll - \beta^{-1}$ (or, equivalently, $\Delta S = \beta W_{\text{diss}} \ll -1$), one can neglect the term $\vert \vert \ket{\Xi_0^\xi} \vert\vert^2 \leq e^{\beta W_\mathrm{diss}}$, and thus obtain the projection $\vert \Psi_W^\xi \rangle_{S,E} \simeq \ket{\Xi_1^\xi}$. In this case, we correspondingly achieve:
\begin{align}
\label{eqn:Xi2}
\mathcal{P}_\xi(W) &= \tilde{P}(-W) \bigl(q^\xi_0 e^{\beta W_\mathrm{diss}}  + q^\xi_1 \bigl) + 2\,\mathbb{Re}\bigl(I_\xi (W)\bigr) \notag \\ &\approx q^\xi_1 ~ \tilde{P}(-W).
\end{align}
Hence, here the joint state is projected onto the time-reversal component of the quantum superposition (as if a projective measurement $\vert 1 \rangle \langle 1\vert_A$ on the auxiliary  system was performed, followed by the observation of the work value $W$). Similarly to the previous case, Eq.~\eqref{eqn:Xi2} provides the probability to get such an outcome in an estimation of the work.


\subsection{Interference effects in the work distribution} \label{sec:interference}

\begin{figure}[bt]
\centering
\includegraphics[width=.95\columnwidth]{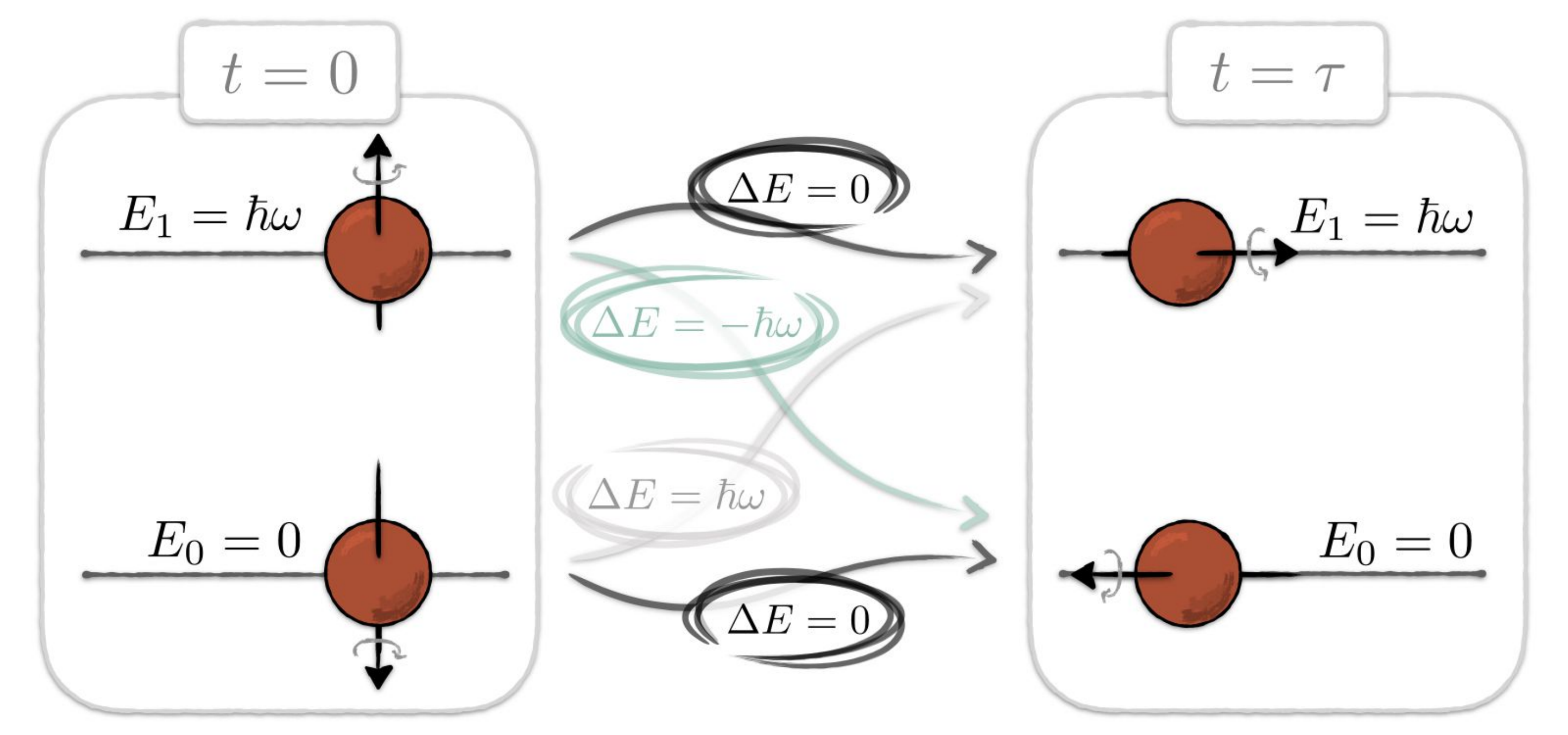}
\captionof{figure}{\footnotesize \textbf{Schematic representation of the two-point measurement scheme in the forward process for our spin-$\frac{1}{2}$ system.} A spin-$\frac{1}{2}$ particle in the thermal state of the initial  Hamiltonian is measured in its eigenbasis $\{\vert z_{\pm} \rangle \}$ at time $t=0$. After the action of the quench described by the time-dependent Hamiltonian in Eq.~\eqref{eqn:Hamiltonian}, it is measured in the eigenbasis $\{\vert x_{\pm} \rangle \}$ of the final Hamiltonian at time $t = \tau$. Depending on the measured states at the two times, the thermodynamic quench causes an energy change $\Delta E = 0, \pm \hbar \omega$, with $\omega$ being the spin's natural frequency, and $\hbar$ the reduced Planck constant.}
\label{img:spin_particle}
\end{figure}

In the previous section we observed that, for individual runs of the process' superposition, whenever the observed entropy production is of the order $\vert \Delta S \vert \gg 1$ (or, equivalently, $\vert W - \Delta F \vert \gg \beta^{-1}$), the system is effectively projected onto a state with a definite thermodynamic time's arrow. 
Conversely, if the measured entropy production is $\vert \Delta S\vert  \lesssim 1$ (or equivalently $\vert W - \Delta F\vert \lesssim  \beta^{-1}$), the superposition state Eq.~\eqref{eqn:M_DSE_Psi0} resulting from the application of the extended TPM scheme lacks a definite time's arrow, exhibiting interference effects.

A closer examination of the term $I_\xi(W)$ highlights the fact that a second source of loss of interference effects in the extended TPM scheme lies in the presence of environmental decoherence, manifested in a negligible overlap between the environmental degrees of freedom, i.e.,  $\braket{\varepsilon_n^{(0)} | \varepsilon_{m}^{(\tau)}} \sim 0$ for all $n,m$. This is the case in all instances where the environment is large and uncontrollable, thus leading the states $\ket{\varepsilon_n^{(0)}}$ and $\ket{\varepsilon_{m}^{(\tau)}}$ to have scarcely any significant overlap. However, for small environments or purposely-engineered environments, such effects can be avoided. 
For instance, one way to implement this scheme would be keeping a sufficiently small path separation in the interferometer in Fig.~\ref{img:sketch_extended}, such that the particle can be assumed to interact with the same environmental degree of freedom regardless of the path it takes. In this specific case, $\braket{\varepsilon_n^{(0)} | \varepsilon_{m}^{(\tau)}} = \delta_{n,m}$.

\begin{figure}[tb]
\centering
\includegraphics[width=.9\columnwidth]{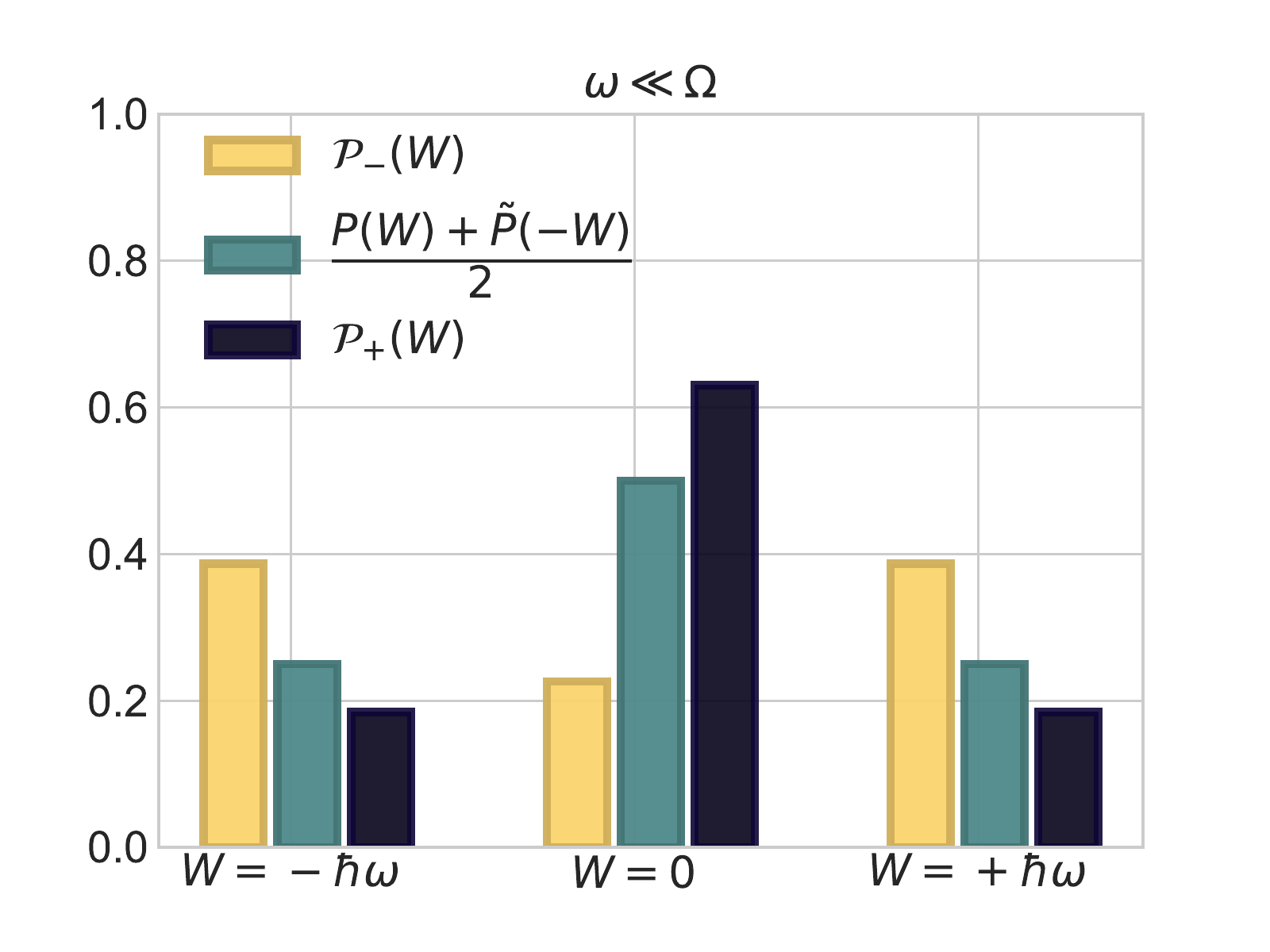}
\captionof{figure}{\footnotesize \textbf{Work probability distribution for a spin-1/2 system undergoing a superposition of forward and its time-reversal thermodynamic process.} The coherent work probabilities $\mathcal{P}_{\pm}(W)$ and the work probabilities of a classical mixture $\bigl(P(W)+\tilde{P}(-W)\bigr)/2$ are compared in the limit of the rapid quench $\omega \ll \Omega$ for $\varphi=\pi$. The results are temperature-independent.} 
\label{img:barplots}
\end{figure}

As an illustrative example, we study the effect of interference in the work distribution in the case of a spin-$\frac{1}{2}$ system, as illustrated in Fig.~\ref{img:spin_particle}. In particular, in the forward quench, the spin system is subjected to a magnetic field whose direction is rotating within the $x - z$ plane at constant angular velocity $\Omega$ around the \textit{y}-axis ($\omega$ being the spin's natural frequency) $H(\Omega t) = \frac{ \hbar \omega}{2} \bigl[\mathbb{1} + \mathrm{cos}\bigl(\Omega t\bigr) \, {\sigma}_z + \mathrm{sin}\bigl(\Omega t\bigr) \, {\sigma}_x\bigr]$. 
In the extended TPM scheme, we superpose the forward quench and its time-reversal twin, and we project the auxiliary  system onto the diagonal basis $\left\lbrace \vert \pm\rangle_A = (\vert 0 \rangle_A \pm \vert 1 \rangle_A)/\sqrt{2} \right\}$. This leads to the work probability distributions $\mathcal{P}_{\pm}(W)$, which illustrates the role played by the interference term. In the limit of a rapid quench ($\omega \ll \Omega$) (and hence of a large degree of irreversibility), the distributions are presented in Fig.~\ref{img:barplots} (yellow and blue bars), together with the one corresponding to a classical mixture of the forward and time-reversal processes (turquoise bars), where here $P(W) = \tilde{P}(-W)$. While the classical mixture displays large fluctuations in the work probability distributions, the contribution of the interference term in $\mathcal{P}_{\pm}(W)$ can sharpen [$\mathcal{P}_{+}(W)$] or flatten [$\mathcal{P}_{-}(W)$] the coherent work distribution, effectively increasing or decreasing the degree of reversibility, respectively. Specifically, the probability that the process will occur in a reversible fashion (i.e., that $W=0$) is higher for $P_+(W=0)$ [lower for $P_-(W=0)$] than for a classical mixture (see Methods-Section~\ref{subsec:spin1/2}). 
In this example, reversibility and adiabaticity coincide, being both reached for slow modulations. In the post-selected case, we can obtain a probability distribution $\mathcal{P}_{+}(W)$ corresponding to that of a slower realisation of the quench. In this sense, through our protocol, one can achieve a net ``speed-up'' of the realisation of an adiabatic quench. 

\section{Discussion}
\label{sec:Discussion}

Viewed in isolation, a thermodynamic system coupled to a reservoir undergoes a dynamic which is generally non-unitary, even though the joint state of the system and the environment evolves in a unitary, reversible fashion. Depending on whether this dynamics favours events involving a positive or a negative change in the total entropy, 
it is possible to establish the temporal direction of the quench which the system has been subjected to (i.e., the time's arrow is aligned along the direction where the total entropy increases~\cite{a_Ftn3}). However, it can be expected that, under some circumstances, the joint state of the system and the environment may as well evolve in an arbitrary superposition of the two, whereby the direction of evolution is controlled by a further quantum system. We note that this superposition of thermodynamic processes does evolve according to an external dynamical time (e.g., the time as shown by the laboratory clock). However, from a quantum-mechanical perspective, there is a-priori no preferential thermodynamic time's arrow (that is, the forward protocol $\Lambda$ and the time-reversal one $\tilde{\Lambda}$ occur in a quantum superposition), and this peculiarity is what this work has explored.
In particular, the core questions behind this work are \textit{i.} how a definite (thermodynamic) arrow of time can emerge in such a picture, and \textit{ii.} what the signature of quantum interference among the forward-in-time and backward-in-time thermodynamic processes is.

We showed that the coherence between the two temporal directions is effectively lost when the entropy production in the process 
is measured: the observation of a large increase (decrease) of dissipative work effectively projects the system in the forward (time-reversal) temporal direction.
It is conceivable to imagine that such a projection could also result from the interaction of the system with the environment, which decoheres the system in a well-defined thermodynamic time's arrow.

On the other hand, for small values of the observed dissipative work (of the order of $\beta ^{-1}$), the system and the auxiliary state may display interference effects. This aspect bears important implications, insofar as, by measuring the state of the control, the system can exhibit a work (entropy production) distribution which is classically impossible with the protocols at hand. This feature can be best observed when both the forward and  the time-reversal processes are, to a high degree, irreversible (i.e., the probability of zero entropy production is low). In this case, indeed, the quantum superposition between the two irreversible processes can result in a dynamics which is no longer such (i.e., the above probability can be significantly increased due to constructive interference). 
Formally, this means that when the distribution of the work $\mathcal{P}_\pm(W)$ is affected by interference effects, this can result in a probability distribution radically different from any classic mixture of $\mathcal{P}(W)$ and $\mathcal{\tilde{P}}(-W)$. As a consequence, 
$\mathcal{P}_\pm(W)$ does not generally satisfy the fluctuation theorem~\eqref{eqn:fluct}.
This is not extremely surprising given that the process generating $\mathcal{P}_\pm(W)$ does not verify the requirements needed for the work fluctuation theorems. In particular, the initial state in Eq.~\eqref{eqn:initial_state} is not a thermal state neither of the system alone, nor of the system together with the control, and the work performed is defined differently in the two quenches of the superposition. Nevertheless, this violation has a crucial implication: it entails that the distribution $\mathcal{P}_\pm(W)$ cannot be generated by any thermodynamic process starting in equilibrium with the environment, and being subsequently driven out of it by means of any given protocol $\Lambda$. Consequently, our procedure provides a recipe to generate thermodynamic processes with a work probability distribution which cannot be reproduced within the 
standard framework of fluctuation theorems.
\section{Methods}

\subsection{Fluctuation theorems and the thermodynamic time's arrow}
\label{subsec:FlTh_TTA}

The link between work fluctuations and the thermodynamic time's arrow can be illustrated in terms of a ``guessing the time directionality game'' which was introduced by C. Jarzynski in Ref.~\cite{doi:10.1146/annurev}. There, the author supposes to record the motion of a non-equilibrium thermodynamic process, and then to toss a coin. Depending on the outcome of the coin, he either plays the movie in the order in which it took place, or in the time-reversal one. In order to determine in which order the movie is being shown, the optimal guessing strategy for a macroscopic system follows from the second law of thermodynamics: if $\langle W \rangle > \Delta F$, the movie proceeds in the correct order, while if $\langle W \rangle < \Delta F$, the movie is being run backwards. Here, $\langle W \rangle$ is the average work performed on the system by the external driving mechanism, and $\Delta F$ the difference in free energies of the thermodynamic states at the beginning and at the end of the movie.
Conversely, for a microscopic system, 
the optimal guessing strategy exploits the so-called ``fluctuation theorems'' \cite{Evans:2002, Esposito:2009, Campisi:2011, Seifert:2012}, together with Bayesian probabilistic reasoning~\cite{Shirts:2003, Maragakis:2008}. We review this study briefly in the Supplementary Note I~\cite{SM}.

In one of its most famous versions~\cite{Bochkov_Kuzovlev_1977,BOCHKOV1981443,Jarzynski:1997,Crooks:1999}, the fluctuation theorem describes the fluctuations of the dissipative work $W_\text{diss}$ associated to the observation of a particular value of $W$ in a single realisation of a non-equilibrium driving protocol 
(i.e., a single shot of the movie):
\begin{equation}
\label{eqn:fluct}
\dfrac{P(+W)}{\tilde{P}(- W)} = e^{\beta W_\text{diss}},
\end{equation}
where $P(+W)$ represents the probability that a work $W$ is invested along the forward thermodynamic evolution, whereas $\tilde{P}(- W)$ is the probability linked to recovering the same amount of work along the time-reversal evolution, both of which start in equilibrium with a thermal bath.
From this equation, it follows that both the probability of total-entropy-decreasing events $(\beta W_\mathrm{diss} < 0)$ in the forward evolution, and that of total-entropy-increasing ones $(\beta W_\mathrm{diss} > 0)$ using the time-reversal dynamics vanish exponentially with the size of the total entropy 
variation:
\begin{subequations}
\begin{align}
&P(\beta W_\mathrm{diss} < -\xi) \leq e^{-\xi}, \\
&\tilde{P}(\beta W_\mathrm{diss} > + \xi) \leq e^{-\xi},
\label{subeq:time-rev}
\end{align}
\end{subequations}
for any $\xi \geq 0$, and where the second inequality \eqref{subeq:time-rev} arises from the fact that, in the time-reversal process, the dissipative work equals $-W_\mathrm{diss}$. In other words, large reductions in the total entropy are unlikely in the forward evolution, while events leading to a large entropy production are unlikely in the time-reversal one. (Notice that the sign of the entropy change is defined to match that of the dissipative work in the forward process.) Interestingly, it is evidenced that, when $\beta W_\text{diss}$ is of the order of one, it is inherently impossible to tell in which of the two orders the process has occurred. In this region, the directionality of time flow cannot be inferred,
and the time's arrow is, so to say, blurred. A clear 
temporal directionality is then reestablished for $\beta \vert W_\text{diss} \vert \gg 1$.

We remark that, here, ``forward'' and ``time-reversal'' are interchangeable labels since each process represents the time-inverted version of the other. Moreover, it is worth noticing that considerations on time-inversion only take on relevance in the absence of complete time-symmetry, as this latter may lead to $\Delta S_\mathrm{tot}$ equal to zero in every single realisation. In order to exhibit time-asymmetry, in the present study the two conjugated processes 
are assumed to start from equilibrium states, a standard procedure in the derivation of fluctuations theorems~\cite{Esposito:2009, Campisi:2011}. This introduces a final (implicit) thermalization step 
which enables irreversibility to emerge (see, e.g., Refs.~\cite{Parrondo:2009,Manzano:2018,Landi:2020}).


\subsection{Case study: a spin-$\frac{1}{2}$ system}
\label{subsec:spin1/2}

In this section, we detail on the interference effects between forward and time-reversal thermodynamic evolution of a spin-$\frac{1}{2}$ system. To this end, we further develop the general expression of Eq.~\eqref{eqn:wdist}. 
Specifically, we project the auxiliary  system onto the diagonal basis $\vert \xi \rangle_A = \left\lbrace \vert \pm\rangle_A = (\vert 0 \rangle_A \pm \vert 1 \rangle_A)/\sqrt{2} \right\}$. This leads to the joint state of the system and the environment $\ket{\Psi_{n,m}^{\pm}}_{S,E,A} \equiv \bigl( \mathbb{1}_{S,E} \otimes |\pm\rangle \langle\pm|_A\bigr) \circ M_{n,m} \vert \Psi_0 \rangle_{S,E,A}$.


The corresponding post-selected work probability distribution, conditioned on the projection of the auxiliary  system onto $\vert \pm \rangle_A$, reads:
\begin{align} \label{eqn:ppn}
\mathcal{P}_{\pm}(W) &= q^\pm_{0} P(W) + q^\pm_{1} \tilde{P}(-W)+ 
2 \, \mathbb{Re}\bigl(I_{\pm}(W)\bigr),
\end{align}
where the interference term $I_{\pm}(W)$ is given by Eq.~\eqref{eqn:Iw} with $\langle 0 | \pm \rangle_A = 1/\sqrt{2}$ and $\langle \pm | 1 \rangle_A =  \pm 1/\sqrt{2}$.
We recall that the states $\Theta \ket{E_n^{(0)}}$ in the above expressions are the eigenstates of the Hamiltonian $\Theta H[\lambda(0)] \Theta^\dagger = H [ \tilde{\lambda}(0)]$. Moreover, we notice that the distribution $\mathcal{P}_\pm(W)$ in Eq.~\eqref{eqn:ppn} differs by the term $I_{\pm}(W) \neq 0$ from what one would have obtained by applying 
the extended TPM scheme to a (classical) convex mixture $|\alpha_0|^2 \, \vert 0 \rangle \langle 0 \vert_A \otimes \rho_0^{\text{th}} + |\alpha_1|^2 \, \vert 1 \rangle \langle 1 \vert_A \otimes \tilde{\rho}_0^{\text{th}}$ of the initial states.


For the outcome $W=0$, the interference term in Eq.~\eqref{eqn:Iw} can be simplified when $\Delta F = 0$, and the sets of eigenvalues of the initial and final Hamiltonians coincide, i.e., $E_n^{(0)} = E_n^{(\tau)}$. 
In that case:
\begin{align}
I_{\pm} (W=0) =& 
 \pm \dfrac{\alpha_0^\ast \alpha_1}{2\, \mathcal{P}(\xi)}~ \sum_{n} p_{n,n} \,e^{-2i \Phi_{n,n}} \notag\\
&\langle \varepsilon_{n}^{(0)} \vert \varepsilon_n^{(\tau)} \rangle \, \bra{E_{n}^{(\tau)}}\Theta \ket{E_n^{(0)}}.
\end{align}
As a result, it emerges that the interference effects can increase (decrease) the probability of observing the work value $W=0$. This yields to a work probability distribution $\mathcal{P}_\pm(W)$ analogous to the one potentially generated by a more reversible (irreversible) process 
than the forward and time-reversal processes themselves, or any classical mixture therefrom. We remark that the interference term $I_\pm(W)$ 
may show non-zero values for $W \neq 0$ in general, as we will see below.

We conclude by evaluating Eq.~\eqref{eqn:ppn} in the concrete example sketched in the main text. We consider a spin system {with natural frequency $\omega$} in a magnetic field $\vec{\lambda}(t)$ whose direction is rotating within the $x - z$ plane at constant angular velocity around the \textit{y}-axis:
\begin{align}
\label{eqn:Hamiltonian}
H\bigl[\vec{\lambda}(t)\bigr] &= \frac{ \hbar \omega}{2} \Bigl[\,\mathbb{1} + \Vec{\lambda}(t) \cdot \Vec{\sigma}\, \Bigr] \notag \\
&= \frac{ \hbar \omega}{2} \Bigl[\,\mathbb{1} + \mathrm{cos}\bigl(\Omega t \bigr) \, {\sigma}_z + \mathrm{sin}\bigl(\Omega t \bigr) \, {\sigma}_x\,\Bigr],
\end{align}
where $\Vec{\lambda}\bigl(t\bigr)=\left(\lambda_0 \, \mathrm{sin}\bigl(\Omega t \bigr), 0, \lambda_0 \, \mathrm{cos}\bigl(\Omega t \bigr)\right)$ and $\lambda_0 = 1$ is the dimensionless magnetic field,
and where the protocol reads $\Lambda = \{\vec{\lambda}(t) ~;~ 0 \leq t \leq  \pi/(2 \Omega) \}$.
We notice that $\Theta H\bigl[\vec{\lambda}(t)\bigr] \Theta^\dagger = H[-\vec{\lambda}(t)]$, implying that the time-reversal of the control parameter corresponds to a flip of the magnetic field.
At the initial and final times of the protocol, the Hamiltonian is diagonal in the $\lbrace \vert z_{\pm} \rangle\rbrace$ and $\lbrace\vert x_{\pm} \rangle\rbrace$ bases, respectively. Therefore, $\vert E_n^{(0)} \rangle = \{ \vert z_{\pm} \rangle_S \}$, with corresponding eigenvalues $E_n^{(0)} = \lbrace 0, \hbar \omega \rbrace$, and $\vert E_m^{(\tau)} \rangle = \{\vert x_{\pm} \rangle_S = \frac{1}{\sqrt{2}}\bigl(\vert z_- \rangle_S \pm \vert z_+ \rangle_S\bigr)\}$, with eigenvalues  $E_m^{(\tau)} = \lbrace 0, \hbar \omega \rbrace$ (we shifted the lower energy level by $\hbar \omega/2$ to avoid negative energy eigenvalues). As a result, $F_0 = F_\tau = - \mathrm{log} \bigl(1 + e^{- \beta \hbar \omega}\bigr)$ and $W_{n,m} = \{\hbar \omega , 0, - \hbar \omega \}$. 

In the frame rotating around the $y$-axis at frequency $\Omega$, the Hamiltonian becomes time-independent, and the unitary governing the evolution can be obtained straightforwardly. Turning back to the Schr\"odinger picture, the applied unitary $U(t,0)$ reads:
\begin{equation}
U(t,0) = e^{- \frac{i}{2} \Omega \sigma_y t} e^{-\frac{i}{2} [\, \omega \, (\mathbb{1} + \sigma_z)  - \Omega \sigma_y \,] \, t}.
\end{equation}
This is used below to compute the work distribution.

\subsubsection{Effect of interference on reversibility}
\label{subsubsec:EffectofInt}

In this subsection, we will represent the environment as a spin-$\frac{1}{2}$ system which is left unaffected during the quench. For instance, we can assume that the purification of the thermal states in Eq.~\eqref{eqn:pur_state_0}-\eqref{eqn:pur_state_tau} read
\begin{subequations}
\begin{align}
& \vert \psi _0 \rangle_{S,E}  = \sqrt{\frac{1}{Z_0}} \, \vert z_- \rangle_S \, \vert z_- \rangle_E + \sqrt{\frac{e^{-\beta \hbar \omega}}{Z_0}} \, \vert z_+ \rangle_S \, \vert z_+ \rangle_E,\\
& \vert \tilde{\psi}_0 \rangle_{S,E}  = \sqrt{\frac{1}{Z_0}} \, \vert x_-  \rangle_S \, \vert z_- \rangle_E + \sqrt{\frac{e^{-\beta \hbar \omega}}{Z_0}} \, \vert x_+  \rangle_S \, \vert z_+ \rangle_E.
\end{align}
\end{subequations}
Furthermore, we will assume to begin the protocol in the state in Eq.~\eqref{eqn:initial_state} with $\alpha_0 = 1/\sqrt{2}$, $\alpha_1 = e^{-i \varphi}/\sqrt{2}$, with $\varphi$ being a controllable phase between the forward and the time-reversal processes.

Next, we compute $\mathcal{P}_\pm(W)$:
\begin{align} \label{eqn:P0}
&\mathcal{P}_\pm(W=0) = \dfrac{1}{2\mathcal{P}(\pm)}\bigl(p_{0,0}+p_{1,1}\bigr) \\
& \mp \dfrac{1}{2\sqrt{2}\mathcal{P}(\pm)} \bigl[\,  p_{0,0} \, \mathrm{cos} \bigl(2 \Phi_{0,0}+\varphi\bigr) + p_{1,1}\,  \mathrm{cos} \bigl(2 \Phi_{1,1}+\varphi\bigr)\bigr], \notag
\end{align}
where we used the fact that $\langle E^{(\tau)}_{n}  \vert \Theta \vert {E}^{(0)}_n \rangle = -1/\sqrt{2}$ for all $n$~\cite{b_Ftn4},
whereas $\langle \varepsilon^{(0)}_{n}  \vert {\varepsilon}^{(\tau)}_n \rangle_E = 1$, and where the marginal probability of the auxiliary system  reads $\mathcal{P}(\pm)=\frac{1}{2}\pm \frac{1}{2\sqrt{2}}\bigl[\,  p_{0,0} \, \mathrm{cos} \bigl(2 \Phi_{0,0}+\varphi\bigr) + p_{1,1}\,  \mathrm{cos} \bigl(2 \Phi_{1,1}+\varphi\bigr)\bigr]$, with $p_{0,0} = \frac{\left \vert \langle x_{-} \vert \, U(\tau, 0)\vert z_{-} \rangle \right \vert^2}{1+e
^{-\beta \hbar \omega}}$, $e^{i \Phi_{0,0}} = \frac{\langle x_{-} \vert \, U(\tau, 0)\vert z_{-} \rangle}{\sqrt{\left \vert \langle x_{-} \vert \, U(\tau, 0)\vert z_{-} \rangle \right \vert}}$, and $p_{1,1} = \frac{\left \vert \langle x_{+} \vert \, U(\tau, 0)\vert z_{+} \rangle \right \vert^2}{1+e
^{-\beta \hbar \omega}}\,e
^{-\beta \hbar \omega}$, $e^{i \Phi_{1,1}} = \frac{\langle x_{+} \vert \, U(\tau, 0)\vert z_{+} \rangle}{\sqrt{\left \vert \langle x_{+} \vert \, U(\tau, 0)\vert z_{+} \rangle \right \vert}}$. From this result, we deduce that it is possible to observe interference between thermodynamic processes occurring in the forward and time-reversal temporal directions.
Following the same procedure for the cases  $W = \pm \hbar  \omega$, we get
\begin{subequations} \label{eqn:Ppm}
\begin{align}
\label{eqn:Pplus}
\mathcal{P}_\pm(W=\hbar \omega) 
= \dfrac{p_{0,1}}{4\mathcal{P}(\pm)}\bigl( \, 1 + e^{-\beta \hbar \omega}\bigr), \\
\label{eqn:Pminus}
\mathcal{P}_\pm(W=-\hbar \omega) 
= \dfrac{p_{1,0}}{4\mathcal{P}(\pm)}\bigl( \, 1 + e^{\beta \hbar \omega}\bigr),
\end{align}
\end{subequations}
which do not feature interference. In the last expressions, $p_{0,1} = \frac{\left \vert \langle x_{+} \vert \, U(\tau, 0)\vert z_{-} \rangle \right \vert^2}{1+e
^{-\beta \hbar \omega}}$, $e^{i\Phi_{0,1}} = \frac{\langle x_{+} \vert \, U(\tau, 0)\vert z_{-} \rangle}{\sqrt{\left \vert \langle x_{+} \vert \, U(\tau, 0)\vert z_{-} \rangle \right \vert}}$, and $p_{1,0} = \frac{\left \vert \langle x_{-} \vert \, U(\tau, 0)\vert z_{+} \rangle \right \vert^2}{1+e^{-\beta \hbar \omega}}\,e
^{-\beta \hbar \omega}$, $e^{i\Phi_{1,0}} = \frac{\langle x_{-} \vert \, U(\tau, 0)\vert z_{+} \rangle}{\sqrt{\left \vert \langle x_{-} \vert \, U(\tau, 0)\vert z_{+} \rangle \right \vert}}$. We illustrate the probability distribution in Eq.~\eqref{eqn:P0}-\eqref{eqn:Ppm} in Fig.~\ref{img:barplots} of the main text.





\subsubsection{Interference terms for varying $\pm \hbar \omega$}

\begin{figure}[tb]
\centering
\includegraphics[width=.95\columnwidth]{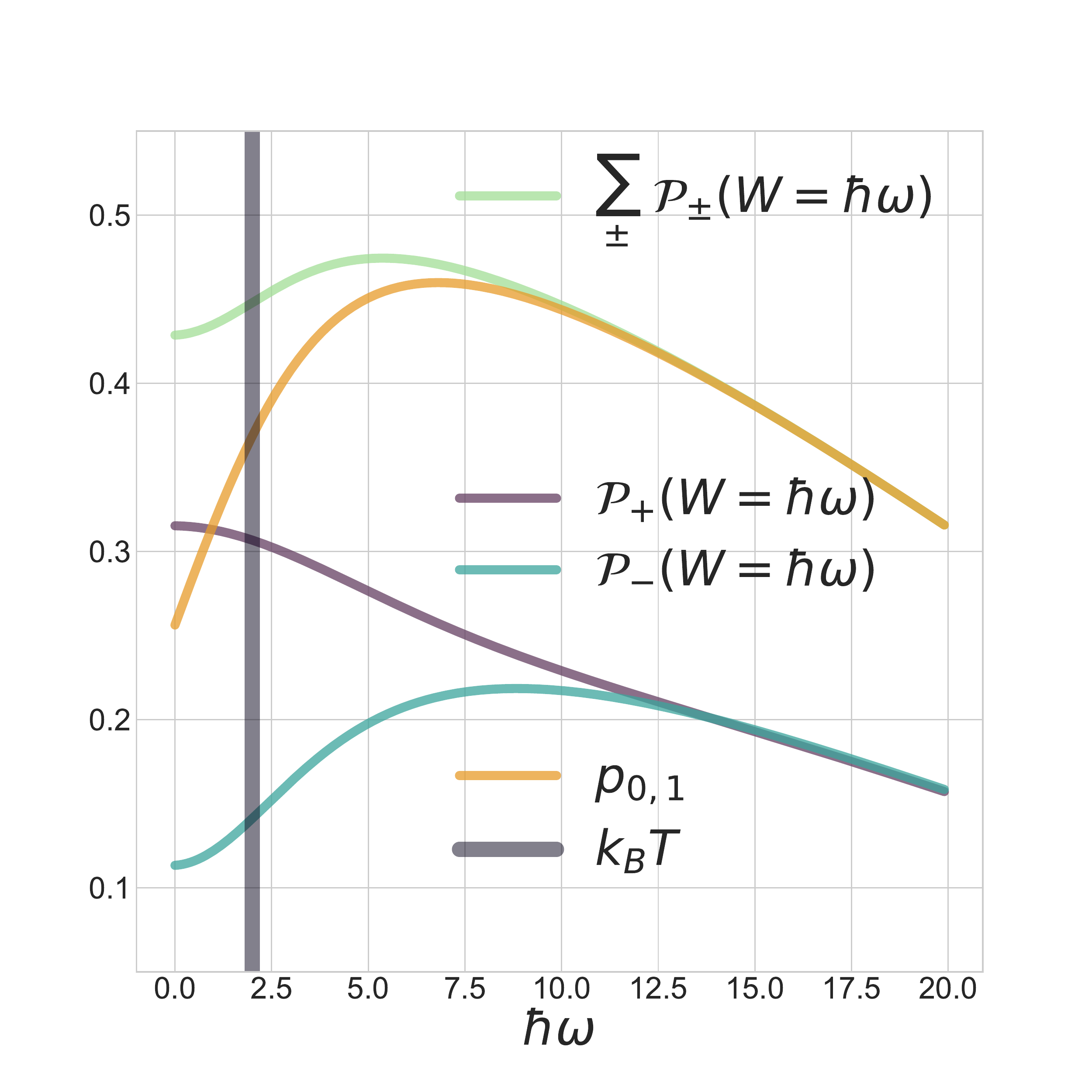}
\captionof{figure}{\footnotesize \textbf{Work probabilities of a spin-1/2 system under the time-dependent Hamiltonian with a varying amount $\hbar\omega$ of work  invested}. For values of $\hbar\omega$ smaller or of the order of $\beta^{-1} = k_B T = 1/2$ ($k_B = \hbar = 1$), the work probabilities $\mathcal{P}_{+}(W = \hbar \omega)$ and $\mathcal{P}_{-}(W = \hbar \omega)$  (see Eq.~\eqref{eqn:pW0pmhbaromegapm}; turquoise and purple curves) strongly depend on the interference terms. For values  $\hbar \omega \gg \beta
^{-1}$, $\mathcal{P}_{+}(W = \hbar \omega)+ $ $\mathcal{P}_{-}(W = \hbar \omega)$ (green curve) tends to the value $p_{0,1}$ (yellow curve), which is obtained by projecting the process to the forward direction and obtaining the work difference $\hbar\omega$. This illustrates that observing large work values $\hbar \omega \gg \beta
^{-1}$ ($\hbar \omega \ll - \beta
^{-1}$) effectively projects the process onto the forward (time-reversal) direction.} 
\label{img:p_W_hbaromega}
\end{figure}

In the previous case study, we represented the environment as a spin-$\frac{1}{2}$ system which is left unmodified by the thermodynamic quench. This caused the cancellation of all interference terms in $\mathcal{P}_\pm(W = \pm\hbar \omega)$. In this subsection, on the contrary, we suppose that the environment undergoes a spin-flip during the quench:
\begin{subequations}
\begin{align}
& \vert \psi _0 \rangle_{S,E}  = \sqrt{\frac{1}{Z_0}} \, \vert z_- \rangle_S \, \vert z_- \rangle_E + \sqrt{\frac{e^{-\beta \hbar \omega}}{Z_0}} \, \vert z_+ \rangle_S \, \vert z_+ \rangle_E,\\
& \vert \tilde{\psi}_0 \rangle_{S,E}  = \sqrt{\frac{1}{Z_0}} \, \vert x_-  \rangle_S \, \vert z_+ \rangle_E + \sqrt{\frac{e^{- \beta \hbar \omega}}{Z_0}} \, \vert x_+  \rangle_S \, \vert z_- \rangle_E.
\end{align}
\end{subequations}
This change results in $\langle \varepsilon^{(0)}_{n}  \vert {\varepsilon}^{(\tau)}_m \rangle_E  = 0$, for $n=m$.
For the sake of simplicity, below we will also set $\varphi=\pi$.

The three probabilities discussed in the previous section become therefore:
\begin{align}
\label{eqn:pW0pmhbaromegapm}
& \mathcal{P}_\pm(W=0) = \dfrac{1}{2 \mathcal{P}(\pm)} \bigl(p_{0,0}+p_{1,1}\bigr), \\
&\mathcal{P}_\pm(W=\hbar \omega) 
= \dfrac{p_{0,1}}{4 \mathcal{P}(\pm)}\Bigl[ \, 1 + e^{-\beta \hbar \omega} \pm \sqrt{2} \, e^{-\frac{ \beta \hbar \omega}{2}} \, \mathrm{cos} \bigl(2 \Phi_{0,1}\bigr)\Bigr], \notag \\
&\mathcal{P}_\pm(W=-\hbar \omega) 
= \dfrac{p_{1,0}}{4 \mathcal{P}(\pm)}\Bigl[ \, 1 + e^{\beta \hbar \omega} \mp \sqrt{2} \, e^{\frac{ \beta \hbar \omega}{2}} \, \mathrm{cos} \bigl(2 \Phi_{1,0}\bigr)\Bigr], \notag
\end{align}
where the marginal probability of the auxiliary system  is now $\mathcal{P}(\pm) = \frac{1}{2}\pm\frac{1}{2\sqrt{2}}\bigl[p_{0,1}\,e^{-\frac{\beta \hbar \omega}{2}} \mathrm{cos}(2\Phi_{0,1})-p_{1,0}\,e^{\frac{\beta \hbar \omega}{2}} \mathrm{cos}(2\Phi_{1,0})\bigr]$, and where $p_{0,0}$, $\Phi_{0,0}$, $p_{1,1}$, and $\Phi_{1,1}$ are the same as in case study \ref{subsubsec:EffectofInt}.

In Fig. \ref{img:p_W_hbaromega}, we show the work probability distributions for varying $\hbar \omega$.
For work values $\hbar \omega$ smaller than, or of the order of $\beta^{-1}$, we observe strong interference effect, as shown by the difference between $\mathcal{P}_{+}(W=\hbar\omega)$ and $\mathcal{P}_{-}(W=\hbar\omega)$. For work values $\hbar \omega \gg \beta^{-1}$, this difference vanishes, and the probability $\mathcal{P}(W = \hbar \omega):=\mathcal{P}_{+}(W=\hbar \omega)+\mathcal{P}_{-}(W=\hbar \omega)$ to obtain the work value $\hbar \omega$ tends to the probability $p_{0,1}$ of first projecting the auxiliary  system onto the forward direction, and then obtaining the work value $\hbar\omega$. This trend shows that the observation of large work values effectively projects the system into a well-defined temporal direction.

\subsection{Other sources of irreversibility}

In this work, we define the orientation of the time's arrow based on the sign of the entropy variation associated to a process consisting of a driving unitary, followed by a thermal relaxation. We identify the ``forward" direction with the one in which entropy increases, and the ``backward" direction with the one in which entropy decreases. Based on this definition, we construct quantum superpositions between opposing time's arrows by superimposing thermodynamic quenches which are time-reversal twins of one another. [This is achieved in spite of an underlying time's arrow pointing ``forward" according to the clock in the laboratory.]
By means of our extended TPM scheme, the superposition is then maintained until the final measurement, whose outcome produces the possible instances of the work exerted on the system (see Fig.~\ref{img:sketch_extended}), and the corresponding entropy production. In this way, we extract a quantity $\Delta S =\beta (W - \Delta F)$, which allows us to deduce whether our system has been projected onto a well-defined temporal axis (whenever $\vert \Delta S \vert \gg 1$), or if we were facing a quantum superposition between opposite time's arrows (when $\vert \Delta S \vert \sim 1$).

A question which may arise is then how one can use the act of measurement ---which itself establishes a well-defined arrow of time--- to conclude that the system has been projected onto either the ``forward'' time direction, the ``backwards'', or even in a quantum superposition of the two. Indeed, the quantum measurement may generate a large entropy production ($\Delta S_{\text{QM}} \gg 1$), and, as such, bears with it a well-defined orientation of the time's axis (that is, the ``forward'' direction). If one were to include this entropy production to the overall entropy computation, investigating the occurrence of superpositions of time's arrow might seem irrelevant since eventually $\Delta S_{\text{QM}} \gg \vert \Delta S\vert$, and the time's arrow would always appear as well-defined (pointing ``forward"). The flaw in this reasoning resides in the fact that, although entropy changes in thermodynamic quenches and entropy changes in quantum measurements both refer to the same figure of merit (and thus can both be used to define an arrow of time), they shall not be added together. The reason is that, when we perform a measurement of the work dissipated in a thermodynamic process (and hence its associated entropy production), the amount of entropy change we read out from the measurement will not include the measurement's entropy (if this were not the case, it would be intrinsically impossible to evaluate only the entropy produced by the thermodynamic quench). In light of this, in this work we focused only on the arrow of time linked to the superposition of thermodynamic quenches. This enabled us to study situations wherein an intrinsically indefinite arrow of time is established despite the presence of other sources of irreversibility (e.g., the laboratory clock, the state preparation, the act of measurement, etc.).


\section*{Data Availability:} All data needed to evaluate the conclusions of the paper are present in the paper and/or the Supplementary Information. Additional data related to this paper will be made available from the authors upon reasonable request.


\renewcommand{\baselinestretch}{1.2}
\bibliographystyle{apsrev4-1}
\bibliography{STA_BIB}

\begin{thebibliography}{52}
\bibitem [{a_F()}]{a_Ftn3}%
  \BibitemOpen
  \href@noop {} {}\bibinfo {note} {Notice that, under our sign convention, this
  means that the time's arrow matches a positive entropy change in the case of
  the forward process, and a negative entropy change for the time-reversal
  process.}\BibitemShut {Stop}%
\bibitem [{b_F()}]{b_Ftn4}%
  \BibitemOpen
  \href@noop {} {}\bibinfo {note} {To evaluate $\langle E^{(\tau)}_{n} \vert
  \Theta \vert {E}^{(0)}_n \rangle_S$, we made use of the fact that the
  time-reversal operator $\Theta$ for a spin-$\frac{1}{2}$ system acts as
  $\Theta = i \sigma_y K$, where $K$ is the complex conjugation operator. Thus,
  $\Theta\vert z_{\pm}\rangle = \mp\vert z_{\mp}\rangle$.}\BibitemShut {Stop}%
\end{thebibliography}

\vspace{2mm}

\section*{Acknowledgments:} The authors wish to thank P. Skrzypczyk for useful comments on the draft, as well as the organisers of the conference ``\textit{New Directions in Quantum Information}'', (Nordita, Stockholm (Sweden); April 1-26 2019) for providing a stimulating platform for the discussion which originated this result. \textbf{Funding:} G.R. acknowledges financial support from the Royal Society through the Newton International
Fellowship No.\ NIF$\backslash \text{R1}\backslash$202512. G.M. acknowledges funding from the European Union's Horizon 2020 research and innovation programme under the Marie Sk\l{}odowska-Curie grant agreement No 801110 and the Austrian Federal Ministry of Education, Science and Research (BMBWF). \v{C}.\ B.\ acknowledges financial support from the Austrian Science Fund (FWF) through BeyondC (F7103-N48), the project no.\ I-2906, from the European Commission via Testing  the  Large-Scale  Limit  of  Quantum  Mechanics (TEQ) (No.\ 766900) project, from Foundational Questions Institute (FQXi) and from the John Templeton Foundation through grant 61466, The Quantum Information Structure of Spacetime (qiss.fr).

\section*{Author contributions:} G.R., G.M. and \v{C}.B. contributed to all aspects of the research, with the leading input of G.R.

\section*{Competing interests:} The authors declare not to have any competing interests.


\onecolumngrid

\newpage
\section*{Supplementary Materials}

\setcounter{equation}{0}
\setcounter{figure}{0}
\setcounter{table}{0}
\makeatletter
\renewcommand{\thesection}{S\arabic{section}}
\renewcommand{\theequation}{S\arabic{equation}}
\renewcommand{\thefigure}{S\arabic{figure}}


\section*{I. Guessing the Time's Direction in a Thermodynamic Process}
\label{appA}

In 1927, Sir A. Eddington introduced the notion of `arrow of time' \cite{Eddington_1928} to refer to the temporal directionality that he saw as deeply rooted in the second law of thermodynamics. He explained that, according to this law, in order to determine the direction in which time is flowing for a macroscopic system subjected to an irreversible process, it is sufficient to examine the relation between the work $W$ performed on the system, and the variation of its free energy $\Delta F$: time must flow in the direction in which $W > \Delta F$. This apparently unequivocal description weakens in the microscopic case, where it is possible to occasionally observe `fluctuations' from the Clausius inequality.  It follows that, in the microscopic case, it is no longer possible to univocally determine the direction of time from the sign of $W - \Delta F$.

With the aim to refine these considerations, in Ref. \cite{doi:10.1146/annurev}, C. Jarzynski evaluated the possibility of defining the temporal direction of a thermodynamic process from a given set of data. In the following, we go over his reasoning briefly.

Let us imagine filming a microscopic system that, subjected to a thermodynamic process $\Lambda(t)$, varies from an initial state at time $t=0$, to a final state at time $t=\tau$. We will suppose that \textit{i.} the camera is able to record the motion of each particle constituting the system, \textit{ii.} we are given full knowledge of the Hamiltonian function of the system $H\bigl(\Lambda(t)\bigr)$, and of the value of $\Delta F = F_\tau-F_0$. Depending on the result obtained from the coin toss, the movie will be shown to us in either the correct or reverse order. Our goal is to determine, based on the given information, whether the movie is shown in the correct or reverse order.

This problem can be addressed using statistical inference. We call $L(F \, \vert \, \gamma)$ the likelihood that the process is shown in the forward direction $F$ if the microscopic trajectory $\gamma$ is shown. Likewise, $L(R \, \vert \, \gamma)$ is the likelihood that the process is shown in the time-reversal direction $R$ given the microscopic trajectory $\gamma$. Obviously, the two terms sum up to one:
\begin{equation}
\label{eqn:likelihood}
L(F \, \vert \, \gamma) + L(R \, \vert \, \gamma) = 1.
\end{equation}
We call $W$ the work performed on the system for the trajectory $\gamma$. For a macroscopic system, according to the Clausius inequality, we have that, if $W> \Delta F$, we are observing the process $F$, whilst we are observing $R$ if $W< \Delta F$. In this case, then, $L(F \, \vert \, \gamma) = \theta \, (W- \Delta F)$, with $\theta(\,\cdot\,)$ being the unity step function. We now evaluate the likelihood corresponding to the microscopic case. From Bayesian theory, we know that
\begin{equation}
L(F \, \vert \, \gamma) = \dfrac{P(\gamma \, \vert \, F) \cdot P(F)}{P(\gamma)},
\end{equation}
where $P(F)$ is the probability that we have been shown the process in the forward direction (\textit{i.e.}, $1/2$), while $P(\gamma) = P(F) \,P(\gamma|F) + P(R) \,P(\gamma|R)$ is a normalization constant. We write the analogous formula for $L(R \, \vert \, \gamma)$, and combine them together in Eq.~\eqref{eqn:likelihood}:
\begin{align}
\label{eqn:likelihood_sum}
 &\dfrac{P(\gamma \, \vert \, F) \cdot P(F)}{P(\gamma)} +  \dfrac{{P}(\gamma \, \vert \, R)  \cdot {P}(R)}{{P}(\gamma)}  =\notag \\
 &=  \dfrac{P(\gamma \, \vert \, F)}{2 P(\gamma)} \bigl[\,1 + e^{-\beta(W - \Delta F)} \,\bigr]= 1.
\end{align}
In addition, we used the fact that ${P}(\gamma \, \vert \, R) = e^{-\beta(W - \Delta F)} \, P(\gamma \, \vert \, F)$, which is one of the main formulations of the fluctuation theorems, and which can be justified as follows:
\begin{align}
\label{eqn:P_ratio}
&\dfrac{P(\gamma \, \vert \, F)}{{P}(\gamma \, \vert \, R)} = \dfrac{e^{-\beta H(\Lambda(0))}}{Z\bigl(\Lambda(\tau)\bigr)} \, \Biggl(\dfrac{e^{-\beta H(\tilde{\Lambda}(0))}}{Z\bigl(\tilde{\Lambda}(\tau)\bigr)} \Biggr)^{-1} \notag \\
&= \dfrac{Z\bigl(\tilde{\Lambda}(\tau)\bigr)}{Z\bigl({\Lambda}(\tau)\bigr)} e^{\beta[H({\Lambda}(\tau)) - H(\Lambda(0))]} = e^{\beta(W - \Delta F)},
\end{align}
where we assumed $H(\tilde{\Lambda}(0))=H(\Lambda(\tau))$, and where, in the second-last equality, we used the fact that, from Liouville's theorem, we know that the volume occupied by the system in the phase space does not change, and therefore $Z\bigl(\tilde{\Lambda}(\tau)\bigr) = Z\bigl({\Lambda}(\tau)\bigr)$. Note that, while this argumentation applies to the case of classical physics, one can arrive at Eq.~\eqref{eqn:P_ratio} also by using the quantum formalism \cite{Campisi:2011}.

From Eq.~\eqref{eqn:likelihood_sum}, we obtain that
\begin{equation}
\label{eqn:fluctuation}
L(F \, \vert \, \gamma) = \Bigl(1 + e^{-\beta(W - \Delta F)}\Bigr)^{-1} = \bigl(1 + e^{-\beta W_{\text{diss}}}\bigr)^{-1},
\end{equation}
where we have called $ W_{\text{diss}} = W - \Delta F$. This equation has been experimentally tested recently in a driven quantum dot setup~\cite{Hofmann:2017}.

\begin{figure}[tb]
\centering
\includegraphics[width=0.8\columnwidth]{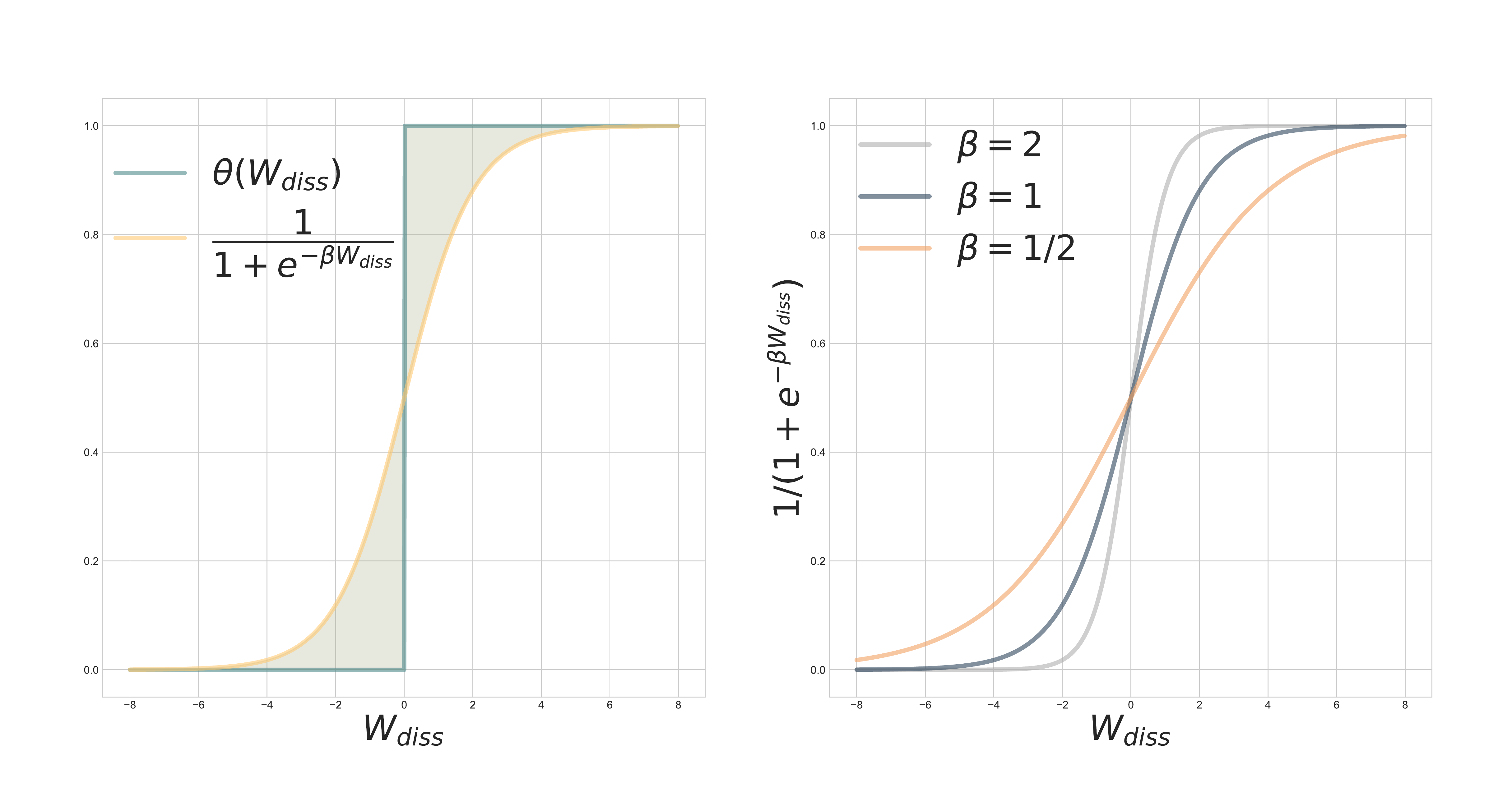}
\captionof{figure}{\footnotesize \textbf{Likelihood $L(F \, \vert \, \gamma)$ as a function of the dissipative work $W_{\text{diss}}$.} {\textbf{(\textit{Left})} \textbf{Comparison between the unity step function (valid in the macroscopic case), and the likelihood $L(F \, \vert \, \gamma)$  (microscopic case).} The shaded areas represent the regions where the time's arrow in the microscopic case is not univocally defined, conversely to the macroscopic one. \textbf{(\textit{Right})}} \textbf{Trend of $L(F \, \vert \, \gamma)$ for different values of the inverse temperature $\beta$.} It is interesting to notice that this function does not depend on the features of the system, nor on the thermodynamic protocol $\Lambda(t)$. In fact, we observe that the lower the inverse temperature (and hence the higher the system's temperature), the wider the region within which the time's arrow is not well-defined.}
\label{img:fluctuation}
\end{figure}

Fig.~\ref{img:fluctuation} shows the discrepancy between the function $\theta(W_{\text{diss}})$, valid in the macroscopic case, and Eq.~\eqref{eqn:fluctuation}, true in the microscopic one. While in the macroscopic case the direction of time is always well-defined, in the microscopic scenario there is a region in which this directionality is genuinely indefinite, and the region widens with increasing the system's temperature.

\section*{II. Description of the External Control Parameter}

It is commonly considered that the unitary $U(t,0)$ acting on the system is implemented by some externally-controlled parameter of the Hamiltonian (classical field). In our case, such an external control could also be included explicitly in the quantum description of the joint state [Eq. (5) of the main text] as an energy reservoir (or `battery') interacting with the system:
\begin{equation}
\label{eqn:initial_state_battery}
\vert \Psi_0 \rangle_{S,E,A,B} = \Bigl(\alpha_0  \, \vert \psi_0 \rangle_{S,E} \otimes \vert 0 \rangle_{A} + \alpha_1  \, \vert {\tilde{\psi}}_0 \rangle_{S,E} \otimes \vert 1 \rangle_{A}\Bigr)\otimes \ket{b}_B,
\end{equation}
where $\ket{b}_B$ corresponds to an arbitrary initial state of the battery system used for the implementation of both quenches $U$ and $\tilde{U}$. It is clear that, in order to allow coherent operations on the system, such a battery needs to be also a source of coherence~\cite{PhysRevLett.113.150402, Malabarba:2015, Korzekwa:2016}. In the limit of the battery acting as an `unbounded' reference frame~\cite{Bartlett:2007} (\textit{i.e.}, infinite source of coherence), a classical-driving is recovered.
More precisely, given an arbitrary unitary $U$ acting on the system alone, one can find an energy-preserving unitary $V(U)$ acting on the enlarged Hilbert space of the system and the battery, such that $\tr_B[V(U) \bigl(\rho_{S} \otimes |b\rangle \langle b|_B\bigr) V(U)^\dagger] \approx  U \rho_{S} U^\dagger$, where $\rho_{S}$ is an arbitrary state of the system (see, \textit{e.g.}, Ref.~\cite{PhysRevLett.113.150402} for a detailed proof). Furthermore, taking the battery to be initially in a strong coherent state, the states of the battery before and after the application of $V(U)$ become almost indistinguishable~\cite{PhysRevLett.113.150402, Dakic2016}. 

For example, following Ref.~\cite{PhysRevLett.113.150402}, the battery may be approximated by a doubly-infinite ladder $H_B = \sum_z \omega \ket{z}\bra{z}_B$ (\textit{e.g.}, an harmonic oscillator far from its ground state), which is assumed to be almost continuous in comparison with any energy spacing in the system, $\omega \ll E_m^{(t)} - E_n^{(t)}$ for all $m,n$ and $t \in [0,\tau]$. Then, arbitrary unitaries on the system may be implemented by inducing rigid translations in the energy ladder. In particular, we consider the global unitary $\mathcal{V} \equiv \ket{0}\bra{0}_A \otimes V(U) + \ket{1}\bra{1}_A \otimes \tilde{V}(\tilde{U})$, which, conditionally on the state of the auxiliary system, applies to each branch in Eq.~\eqref{eqn:initial_state_battery} the corresponding system-battery unitaries:
\begin{subequations}
\begin{align}
V(U) &\equiv \sum_{n,m} \ket{E_m^{(\tau)}}\bra{E_m^{(\tau)}} U(\tau,0) \ket{E_n^{(0)}}\bra{E_n^{(0)}} \notag\\
&\qquad \otimes \Delta(W_{n, m}/\omega), \\
\tilde{V}(\tilde{U}) &\equiv \sum_{n,m} \Theta \ket{E_n^{(0)}}\bra{E_n^{(0)}} \Theta^\dagger \tilde{U}(\tau, 0) \Theta \ket{E_m^{(\tau)}}\bra{E_m^{(\tau)}} \Theta^\dagger \notag\\
&\qquad \otimes \Delta(-W_{n, m}/\omega),    
\end{align}
\end{subequations}
where we introduced the battery translation operator $\Delta(\delta) \equiv \sum_k \ket{k + \delta }\bra{k}_B$, verifying $\Delta(\delta)^\dagger = \Delta(-\delta)$ and $\Delta(\delta_1) \Delta(\delta_2) = \Delta(\delta_1 + \delta_2)$. Notice that, in $V(U)$, any energetic transition $\ket{E_n^{(0)}} \rightarrow \ket{E_m^{(\tau)}}$ induced by $U$ on the system is exactly compensated by a translation on the battery of proportional magnitude, $\Delta(W_{n, m}/\omega)$, where we recall that $W_{n,m} = E_m^{(\tau)} - E_n^{(0)}$. Similarly, in $\tilde{V}(\tilde{U})$ the time-reversed transitions $\Theta \ket{E_m^{(\tau)}} \rightarrow \Theta \ket{E_n^{(0)}}$ are compensated by the opposite battery translations. As a consequence, the global Hamiltonian of the system, the auxiliary qubit and the battery at initial and final times, $\mathcal{H}_0 \equiv \ket{0}\bra{0}_A \otimes (H[\lambda(0)] + H_B) + \ket{1}\bra{1}_A \otimes (\tilde{H}[\lambda(\tau)] + H_B)$ and $\mathcal{H}_\tau \equiv \ket{0}\bra{0}_A \otimes (H[\lambda(\tau)] + H_B) + \ket{1}\bra{1}_A \otimes (\tilde{H}[\lambda(0)] + H_B)$, generate exactly the same energy distribution when applied, respectively, to the global initial and final states $\vert \Psi_0 \rangle_{S,E,A,B}$ and $\mathcal{V} \vert \Psi_0 \rangle_{S,E,A,B}$ (assuming the auxiliary qubit internal states $\ket{0}$ and $\ket{1}$ to have the same energy). This guarantees energy conservation.

In this situation, it is convenient to consider that the battery is initiated in a coherent state of the form $\ket{\eta(L,l_0)}_B = \sum_{l=0}^{L-1} \ket{l+l_0}_B/\sqrt{L}$, corresponding to a highly coherent state of length $L$. These states verify~\cite{PhysRevLett.113.150402}: 
\begin{equation}
\langle \eta(L,l_0)| \Delta(\delta) | \eta(L,l_0) \rangle = \max(0, 1-|\delta|/L),    
\end{equation}
and hence the displaced state $\Delta(\delta) | \eta(L,l_0) \rangle$ becomes indistinguishable from the original state $\ket{\eta(L,l_0)}$ when $L \gg |\delta|$. Therefore, whenever we choose in Eq.~(\ref{eqn:initial_state_battery}) the initial state of the battery as $\ket{b}_B \equiv \ket{\eta(L,l_0)}$ for $L \gg \max_{n,m}(|W_{n, m}|/\omega)$, the back-reaction over the battery due to the implementation of the quenches $U$ and $\tilde{U}$ may be safely neglected. This implies that, in any such protocol and for any outcomes $m$ and $n$ of the extended two-point measurement scheme introduced in Sec.~II, the associated changes in the state of the battery would be unnoticeable.

In light of this, assuming the battery to be in a strong coherent state in the amplitude corresponding to both the forward and the time-reversal directions, we conclude that the battery ends in a nearly-indistinguishable state from its initial one, and that it can be, to a good approximation, factorized from the rest. Consequently, considering explicitly the battery in the quantum description of the joint state does not introduce any extra source of decoherence in our interferometric scheme, and hence the battery can be fully replaced by a classical external control.

Finally, we also remark that in our interferometric setup we do not require the battery to be used `catalytically'~\cite{PhysRevLett.113.150402}, hence avoiding the accumulation of (finite-size) errors leading to the battery degradation~\cite{Vaccaro:2018}. For instance, the battery could be reprepared in its ready-to-work state at the beginning of every realization of our scheme.

\section*{III. Relation between Entropy Production and Work}

In the main body, all our results are formulated in terms of the work performed during the quench. There is, however, a link between this latter and the entropy production~\cite{Kawai:2007,Parrondo:2009,Batalhao:2015}. Indeed, the \textit{stochastic entropy production} can be constructed from the stochastic work as:
\begin{equation}
\label{eqn:stochastic_entropy}
\Delta S_{n,m} := \beta \bigl(W_{n,m} - \Delta F \bigr),
\end{equation}
where $\Delta F := F_\tau - F_0 = -\log(Z_\tau/ Z_0)$ is the difference in free energies between the equilibrium states at times $t = \{0, \tau \}$. Again, as a consequence of $\tilde{p}_{n\vert m} = p_{m\vert n}$, a generalized version of the fluctuation theorem for the stochastic entropy production in Eq.~\eqref{eqn:stochastic_entropy} can be obtained~\cite{Kawai:2007, doi:10.1142/9789814425193_0003}:
\begin{equation} \label{eqn:fluct2}
\mathrm{ln} \Bigl(\dfrac{p_{n,m}}{\tilde{p}_{m,n} }\Bigr) = \mathrm{ln} \Biggl(\dfrac{p_n^{(0)}}{\tilde{p}_m^{(0)}}\Biggr) = \Delta S_{n,m}.
\end{equation}
This equation conveys a well-defined meaning to the entropy production in terms of irreversibility by linking it to the ratio between the probability of transitions $\ket{E_n^{(0)}} \rightarrow \ket{E_n^{(\tau)}}$ in the forward dynamics, and the probability of the inverse transition $\Theta \ket{E_m^{(\tau)}} \rightarrow \Theta \ket{E_n^{(0)}}$ in the time-reversal dynamics.
Moreover, following Eq.~\eqref{eqn:fluct2}, reversible processes, for which $p_{n,m} = \tilde{p}_{m,n}$, necessarily produce zero entropy for every single realization of the protocol $\Lambda$, \textit{i.e.}, $\Delta S_{n,m} = 0$ (or, equivalently, $W_{n,m} = \Delta F$) for all $n,m$.

\end{document}